\documentclass[12pt, draftclsnofoot, onecolumn]{IEEEtran}
\usepackage{graphicx,amsmath,amssymb}
\usepackage{subfigure}
\usepackage{citesort}
\usepackage{fancyhdr}
\usepackage{mdwmath}
\usepackage{mdwtab}
\usepackage{balance}
\usepackage{xcolor}
\usepackage{bm}
\usepackage{amsthm}
\usepackage{algorithm}
\usepackage{algorithmic}
\usepackage{multirow}
\usepackage{flafter}
\usepackage{setspace}
\usepackage{cite}

\newtheorem{remark}{Remark}
\newtheorem{proposition}{Proposition}
\newtheorem{theorem}{Theorem}

\newtheorem{lemma}{Lemma}

\newtheorem{corollary}{Corollary}

\newtheorem{assumption}{Assumption}

\begin{document}
\title{Modeling and Coverage Analysis for RIS-aided NOMA Transmissions in Heterogeneous Networks}

\author{Ziyi~Xie,~\IEEEmembership{Student Member,~IEEE,} Wenqiang~Yi,~\IEEEmembership{Member,~IEEE,} Xuanli~Wu,~\IEEEmembership{Member,~IEEE,} Yuanwei~Liu,~\IEEEmembership{Senior Member,~IEEE,} and Arumugam~Nallanathan,~\IEEEmembership{Fellow,~IEEE}
\thanks{Z. Xie and X. Wu are with the School of Electronics and Information Engineering, Harbin Institute of Technology, Harbin 150001, China (email: \{ziyi.xie, xlwu2002\}@hit.edu.cn).}
\thanks{W. Yi, Y. Liu, and A. Nallanathan are with the School of Electronic Engineering and Computer Science, Queen Mary University of London, E1 4NS, U.K. (email: \{w.yi, yuanwei.liu, a.nallanathan\}@qmul.ac.uk).}
\thanks{Part of this work was submitted to IEEE Global Communications Conference (GLOBECOM), 2021 \cite{0conference}.}
}

\maketitle
\vspace{-2 cm}
\begin{abstract}
\vspace{-0.5 cm}
  Reconfigurable intelligent surface (RIS) has been regarded as a promising tool to strengthen the quality of signal transmissions in non-orthogonal multiple access (NOMA) networks. This article introduces a heterogeneous network (HetNet) structure into RIS-aided NOMA multi-cell networks. A practical user equipment (UE) association scheme for maximizing the average received power is adopted. To evaluate system performance, we provide a stochastic geometry based analytical framework, where the locations of RISs, base stations (BSs), and UEs are modeled as homogeneous Poisson point processes (PPPs). Based on this framework, we first derive the closed-form probability density function (PDF) to characterize the distribution of the reflective links created by RISs. Then, both the exact expressions and upper/lower bounds of UE association probability are calculated. Lastly, the analytical expressions of the signal-to-interference-plus-noise-ratio (SINR) and rate coverage probability are deduced. Additionally, to investigate the impact of RISs on system coverage, the asymptotic expressions of two coverage probabilities are derived. The theoretical results show that RIS length is not the decisive factor for coverage improvement. Numerical results demonstrate that the proposed RIS HetNet structure brings significant enhancement in rate coverage. Moreover, there exists an optimal combination of RISs and BSs deployment densities to maximize coverage probability.
\end{abstract}
\vspace{-0.5 cm}
\begin{IEEEkeywords}
\vspace{-0.5 cm}
 Heterogeneous network, non-orthogonal multiple access, reconfigurable intelligent surface, stochastic geometry, user association
\end{IEEEkeywords}
\vspace{-0.5 cm}

\vspace{-0.5 cm}
\section{Introduction}

To enhance energy efficiency and network throughput, the controllable wireless communication environment becomes a new requirement for future networks. Reconfigurable intelligent surface (RIS), also called intelligent reflecting surface (IRS), has been envisioned as a promising tool for providing this ability \cite{0surveyLiu,0surveyZhang,0EE}. Specifically, RIS uses metamaterial or patch-array based technologies to alter the amplitude and/or phase of signals reflected by it \cite{0surveyLiu,0RISmaterial}. Compared with conventional active relays and small base stations (BSs), RISs are nearly passive and the energy is mainly consumed during the control and programming phase \cite{0surveyMarco}. Based on the capability of smartly manipulating signal propagation, there are a variety of protential applications of RISs. One promising application is that RISs can be deployed to create strong line-of-sight (LoS) reflective links for blocked user equipment (UE). This coverage enhancement is of great significance for wireless networks, especially those with higher frequency bands such as millimeter wave and terahertz system, whose communication is vulnerable to blockages \cite{0relaycomp,1blockage}.

For further enhancing the spectral efficiency of RIS-aided wireless networks, non-orthogonal multiple access (NOMA) has been considered to be an efficient booster as it is able to exploit the power domain to allocate the same resource block to multiple UEs \cite{0NOMA1,0NOMA2,0NOMA3}. Leveraging the difference of channel conditions among UEs, NOMA obtains the channel capacity gain over traditional orthogonal multiple access (OMA) schemes when considering UE fairness \cite{0roleNOMA}. NOMA systems also benefit from the deployment of RISs. For NOMA UEs with weak channel gains, RISs are capable of creating stronger transmission paths. Moreover, since RISs are able to change the channel conditions of UEs, RISs offer a flexible decoding order for NOMA with different QoS requirements \cite{1rismodel}.

\subsection{Related Works}
For RIS-enabled communications, recent works have paid significant attention to system designs and performance analysis. As mentioned previously, RISs can be applied in various scenarios due to their nature of creating reflective paths. In \cite{1indoor}, a three dimensional (3D) indoor communication model was considered and several RISs coated on walls were used to assist signal transmissions whose direct links are blocked. Regarding outdoor models, an IRS was located at a cell edge to improve the communication quality of a nearby UE in \cite{1celledge}. In \cite{1terahertz}, a holographic RIS was deployed to provide passive beamforming between the servig BS and UEs under terahertz systems. The authors in \cite{1airground} employed IRSs for air-ground networks to control the reflection direction of ratio waves and hence the inter-cell interference can be suppressed. These works are under the assumption that the locations of RISs are known and the transmission paths are predefined. To characterize the random locations of RISs in large-scale deployment networks, stochastic geometry is an efficient mathematical tool. A recent work \cite{1singlecell} focused on spatial thoughput in a multi-user single-cell network, where IRSs are randomly deployed in a ring area. Average performance for multi-cell networks was evaluated in some research articles, such as \cite{1blockage,1LIS,1gammaapproxi}, while the complex reflective channels impose challenges on UE association and subsequent analysis. In \cite{1blockage}, the authors studied a special case that UEs in blind-spot areas obtain indirect LoS links through RISs coated blokages. For tractability, the authors in \cite{1LIS} proposed a two-step association strategy in a millimeter wave multi-antenna system. Based on this strategy, the enhancement of capacity and energy efficiency is validated. In \cite{1gammaapproxi}, UEs associated to their nearest BSs and the corresponding communication was also assisted by IRSs. This work showed that the composite channel gain can be approximated by the Gamma distribution.

Sparked by the aforementioned potential benefits of integrating RISs and NOMA, the majority of research efforts have been devoted to RIS-NOMA networks. In \cite{2NOMAdesign}, the authors proposed a design that an IRS was deployed to assist the transmission from a BS to cell-edge NOMA users while cell-center NOMA users associated to the BS directly. The authors in \cite{2deadzone} focused on the dead zone users and jointly optimized the beamforming vectors at BSs and IRSs. The above RIS-aided paradigms assumed that there is no direct links between users and BSs. In practice, BS-UE links always exist although the channel gain may be degraded by blockages, so in \cite{2NOMAcomposite,2RA1,2RA2}, both BS-UE links and BS-RIS-UE links were considered. In \cite{2NOMAcomposite}, the authors assumed all RIS-related channels are LoS while BS-user channels are non-line-of-sight (NLoS), and presented an optimization framework to minimize the total transmit power. Two insightful works, \cite{2RA1} and \cite{2RA2}, solved the joint optimization problem over multiple factors including RIS parameters, subchannel assignment, and decoding order for maximizing system sum rates.
Furthermore, some works have contributed to the analysis of theoretical performance in RIS-NOMA networks. The authors in \cite{2NOMAcomp} compared the outage probability and ergodic rate between NOMA and OMA in IRS-enabled communications and validated the enhancement from IRS-NOMA schemes. The authors in \cite{2NOMAsingle1} evaluated the network performance by deriving the best-case and worst-case of channel statistics. Stochastic geometry models were considered in \cite{2NOMAsingle1,2NOMAsingle2,1rismodel}. The authors in \cite{2NOMAsingle1} and \cite{2NOMAsingle2} modeled the locations of users as a homogeneous Poisson point processes (PPPs) and studied the spatial effect in single-cell networks. The authors in \cite{1rismodel} introduced a tractable path loss model for large-scale deployment of RISs and extended the RIS-aided model to multi-cell scenarios.

\subsection{Motivations and Contributions}
As discussed in the previous section, RIS is an efficient solution for the performance enhancement of NOMA systems. Although the BS-RIS-UE link significantly strengthens the quality of signal transmissions, most existing works assumed the direct links between BS and UE are weak or blocked but neglected the probability of LoS direct links. In these works, signal transmissions were always assisted by RISs hence UE association schemes were scarcely investigated. Furthermore, due to the complexity of the reflective paths provided by RISs, the research on theoretical performance in multi-cell scenarios is still insufficient.

Motivated by these three seasons, in this paper, we introduce a heterogeneous network (HetNet) structure into RIS-aided NOMA multi-cell networks.
A UE association scheme is provided according to channel conditions.
Considering the deployment of devices in large-scale RIS-NOMA networks, we use stochastic geometry for system modeling and performance evaluation.
The main contributions are summarized as follows:
\begin{itemize}
	\item We provide an analytical framework for the downlink NOMA RIS HetNets based on stochastic geometry. In this framework, We model the random locations of RISs, BSs, and UEs by independent homogeneous PPPs. To maximize channel gains for signal transmissions, we employ a practical UE association rule in which UEs associate either a BS or RIS with higher average received power, where correlated RIS channels are
	considered. Moreover, we consider LoS/NLoS transmissions for the direct BS-UE links and LoS transmissions for the BS-RIS-UE links to depict a general scenario.
	\item We characterize the distance distribution of BS-RIS-UE links and derive its closed-form probability density function (PDF). Based on general path models for BS-UE links and BS-RIS-UE links, we derive the UE association probability of a typical UE in paired NOMA UEs. We also provide the bounds of the UE association probability and show the impact of LoS probability.
	\item We derive the analytical expressions for the signal-to-interference-plus-noise-ratio (SINR) and rate coverage probability of the typical NOMA UE. To investigate the impact of RISs, we deduce the closed-form asymptotic expressions of these two performance metrics versus the half-length of RISs $L$. The theoretical results indicate that we cannot obtain the maximum achievable performance only by enlarging the length of RISs. Additionally, a tractable approximation is calculated for the rate coverage.
	\item The simulation results validate our theoretical analysis and show that 1) compared with OMA or non-RIS scenarios, the proposed NOMA HetNet structure enhances the coverage probability and thus improving the spectral efficiency; 2) the densification of BSs enhances signal transmissions for both BS-UE links and BS-RIS-UE links; 3) there exists an optimal RIS length to maximize the system performance; 4) when the density of BSs is predefined, we can choose an optimal RISs deployment density to maximize the system performance.
\end{itemize}

\subsection{Organizations and Notations}
The remainder of this paper is organized as follows. In section II, we decribe the system model of NOMA RIS HetNets. In section III, we present the path loss models including angle and distance distributions. Additionally, we derive the association probability for each tier.
In section IV, we derive the analytical expressions of SINR and rate coverage probability. In section V, we illustrate numerical results. In section VI, we propose our conclusion. Notations in this paper are listed in Table \ref{table: notations}.

\begin{table*}
	\centering
	\caption{Table of Notations}\label{table: notations}	
	\begin{tabular}{m{2.3cm}<{\centering} |m{12.5cm}<{\centering}}
		\hline \hline
		\textbf{Notation} & \textbf{Description} \\ \hline
		$\Phi_B$;$\Phi_R$;$\Phi_U$ & PPP of BSs; PPP of RISs; PPP of UEs \\ \hline
		$\lambda_B$;$\lambda_R$;$\lambda_U$ & Density of BSs; density of RISs; density of UEs \\ \hline
		${\tilde \Phi}_B$;${\tilde \lambda}_B$ & PPP of active BSs; density of active BSs \\ \hline
		$P_B$;$W$ & Transmit power of BSs; system bandwidth \\ \hline
		$C_ \kappa$;$C_R$ & Intercept of BS-UE links ($\kappa \in \{L,N\}$); intercept of BS-RIS-UE links\\ \hline
		$\alpha_ \kappa$;$\alpha_R$ & Path loss exponent of BS-UE links ($\kappa \in \{L,N\}$); path loss exponent of BS-RIS-UE links\\ \hline
		$m_\kappa$;$m_R$ & Nakagami coefficient of BS-UE links ($\kappa \in \{L,N\}$); Nakagami coefficient of BS-RIS-UE links \\ \hline
		$\theta_{BR}$;$\theta_{RU}$;$L$ & Angle of arrival; angle of departure; half-length of RIS \\ \hline
		$a_s$;$a_l$ & Power allocation factor of small path loss UE;power allocation factor of large path loss UE \\ \hline
		$\beta$; $\sigma^2$ & Blockage parameter; thermal noise\\ \hline \hline
	\end{tabular}
\end{table*}
\section{System Model}

This work considers RIS-aided downlink NOMA networks with a HetNet structure, where BSs, RISs, and UEs are modeled as three independent homogeneous PPPs ${\Phi _B}$, ${\Phi _R}$, and ${\Phi _U}$ in $\mathbb{R}^2$ with density $\lambda _B$, $\lambda _R$, and $\lambda _U$, respectively. The transmit power of BSs is $P_B$ and the system bandwidth is $W$. Since a BS is silent if there is no UE associated to it, the locations of active BSs also obeys PPP with density ${\tilde \lambda _B} = {\lambda _B}\left( {1 - {{\left( {1 + \frac{{{\lambda _U}}}{{3.5{\lambda _B}}}} \right)}^{ - 3.5}}} \right)$, \cite{3actBS}. Both BSs and UEs are equipped with a single antenna. For RISs, we consider a linear model with 2$L$ in length \cite{3rismodelcon}. According to NOMA principles, two NOMA UEs are grouped in each resource block (RB) to improve the spectral efficiency. For simplicity, we assume that one of the paired UEs has been associated to a BS in the previous UE association process, and hence its communication distance $d_C$ is known at the serving BS. A typical UE is randomly selected from $\Phi_U$, which joints the same RB of one connected user to form a NOMA group. The location of this typical UE is fixed at the origin of the considered plane.

\subsection{Channel Model}
In this RIS HetNet, there are two types of communication links between BSs and UEs: 1) BS-UE link, the link that a BS transmit signals directly to its served UEs; 2) BS-RIS-UE link, the link that an assisted RIS is used to reflect signals from a BS to its UEs. In the former case, we adopt a stochastic blockage model for LoS/NLoS propagation \cite{2losmodel}. The blockages are modeled as a rectangle Boolean scheme and the LoS transmission probability between BSs and UEs is shown as
\begin{align}\label{LoS probability}
    {p_{\rm{L}}}(d_0) = {e^{ - \beta d_0}},
\end{align}
where $\beta$ is a parameter determined by the density and the average size of the blockages.

\begin{figure*} [t!]
\centering
\includegraphics[width= 6.5in] {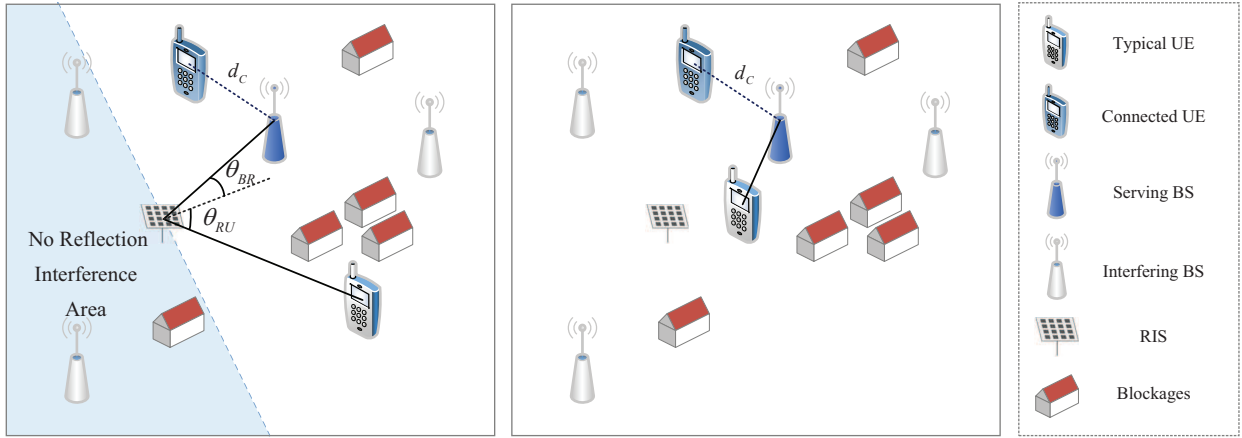}
 \caption{Illustration of the UE association model: (a) Left: The typical UE associates to BS through BS-RIS-UE link; (b) Right: The typical UE associates to BS through BS-UE link.}
\label{system_model}
\end{figure*}

For the BS-UE links, the path loss can be expressed as
\begin{align}
    {L_{BU}}(d_0) = \sum\limits_{\kappa  \in {\rm{\{ L,N\} }}} {\mathbb{B}\left( {{p_\kappa }\left( {{d_0}} \right)} \right)} {L_{BU,\kappa} }\left( {{d_0}} \right),
\end{align}
with
\begin{align}
    {L_{BU,\kappa}}(d_0) = C_\kappa d_0^{-\alpha_\kappa},
\end{align}
where $C_ \kappa$ is the path loss at a reference distance $d_0=1$ and $\alpha_ \kappa$ denotes the path loss exponent. The $\kappa = L$ and $\kappa =N$ represent LoS and NLoS links, respectively. The probability for NLoS transmossions is $p_N(d_0)=1-p_L(d_0)$. $\mathbb{B}(t)$ is a Bernoulli random variable with a probability of success $t$.

For the BS-RIS-UE links, the angle of arrival at a RIS is denoted by $\theta _{BR}$ and the angle of departure is denoted by $\theta _{RU}$. According to \cite{1rismodel}, the path loss under correlated channels can be expressed as
\begin{align}\label{model: RIS path loss}
   {L_{RIS}}({d_{BR}},{d_{RU}}) = {C_R}{\left( {d_{BR}}{d_{RU}} \right)^{ - {\alpha _R}}},
\end{align}
where ${d_{BR}}$ denotes the distance between BS and RIS. The ${d_{RU}}$ is the distance between RIS and the typical UE. The intercept is denoted by ${C_R} = \frac{L^2}{{16\pi
^2}}\left( {\cos (\theta _{BR}) + \cos (\theta _{RU})} \right)^2$. The $\alpha _R$ is the path loss exponent. It should be noted that the BS-RIS-UE links exist only when the BSs and the UEs are distributed at the same side of the RIS.

\subsection{UE Association in RIS HetNets}\label{model: UE Association}
For the UE association, we assume that the typical UE connects to the BS or RIS with the highest received power. In other words, the typical UE associates to the nearest LoS BS or associates to the nearest RIS which reflects signals from its nearest BS. Let $\Phi _U^L$ and $\Phi _U^R$ denote the set of UEs associated to LoS BSs and the set of UEs associated to BSs with the aid of RISs, respectively. Therefore, we have $\Phi _U^L \cup \Phi _U^R = \Phi _U$.

If the typical UE $u_0 \in \Phi _U^L$ associates to the BS $j \in \Phi _B$, the average received power of the desired signal at the typical UE can be expressed as
\begin{align}
    {P_{0,j}} = {a_t}{P_B}L_{BU,L}(d_{0,j}),
\end{align}
where $d_{0,j}$ is the distance between the typical UE and its nearest LoS BS. The $a_t$ is the power allocation factor for the typical UE.

On the other hand, if the typical UE $u_0 \in \Phi _U^R$ associates to the BS $j \in \Phi _B$ through the RIS $i \in \Phi _R$, the average received power of the desired signal at the typical UE can be expressed as
\begin{align}
    P_{0,j}^{(i)} = {a_t}{P_B}L_{RIS}(d_{BR,j}^{(i)}, d_{RU,0}^{(i)}),
\end{align}
where $d_{BR,j}^{(i)}$ is the distance between the serving BS and the assistant RIS. The $d_{RU,0}^{(i)}$ is the distance between the assistant RIS and the typical UE.

For the connected UE, it associates to the BS through BS-UE link, so the average received power of the desired signal at the connected UE can be expressed as
\begin{align}
   {P_C} = {a_c}{P_B}L_{BU,L}(d_C).
\end{align}
where $a_c$ is the power allocation factor for the connected UE.

\subsection{SINR Analysis}
Since the average performance of wireless communications mainly depends on the path loss, we assume the successive interference cancelation (SIC) in a NOMA group is processed at the UE with the smaller path loss. As it is not pre-determined whether the path loss of the typical UE is large or small, we have the following two cases.
\subsubsection{Small Path Loss Case}
When the typical UE has the smaller path loss than the connected UE, the typical UE
first decodes the information of the connected UE associated to the same BS. For power allocation, $a_l$ and $a_s$ are the power allocation factors for large path loss UE and small path loss UE, respectively. The power allocation factors also satisfy the conditions that $a_s \le a_l$ and $a_l+a_s=1$. Thus, in this case, $a_t = a_s$ and $a_c = a_l$. For simplicity, we denote ${\tilde c_{LR}} = \frac{{{C_L}}}{{{C_R}}}$, ${\tilde c_{RL}} = \frac{{{C_R}}}{{{C_L}}}$, ${\tilde \alpha _{LR}} = \frac{{{\alpha _L}}}{{{\alpha _R}}}$ and ${\tilde \alpha _{RL}} = \frac{{{\alpha _R}}}{{{\alpha _L}}}$ in the following parts of this work.

If the typical UE $u_0 \in \Phi _U^L$, ${d_{0,j}} \le {d_C}$ holds and the SINR for the SIC process at $u_0$ is given by
\begin{align}
    \gamma _{t \to c,small}^L = \frac{{{a_l}{P_B}{L_{BU,L}}(d_{0,j}){h_{0,j}^L}^2}}{{{a_s}{P_B}{L_{BU,L}}(d_{0,j}){h_{0,j}^L}^2 + {I_L} + {I_N} + {\sigma ^2}}},
\end{align}
where ${I_L} = \sum\nolimits_{k \in \Phi _B^L\backslash j} {{P_B}{L_{BU,L}}(d_{0,k}){h_{0,k}^L}^2} $ is the interference from other LoS BSs. The ${I_N} = \sum\nolimits_{k \in \Phi _B^N} {{P_B}{L_{BU,N}}(d_{0,k}){h_{0,k}^{N}}^2} $ is the interference from NLoS BSs. For $\kappa =\{L,N\}$, ${h_{0,k}^\kappa}^2$ is the small scale fading power from the BS $k \in \Phi_B$. We characterize the small scale fading as Nakagami-$m$ distribution with an integer parameter $m_\kappa$. Moreover, $\sigma ^2$ is the additive white Gaussian noise power.

After the SIC process, the decoding SINR at the typical UE $u_0$ can be expressed as
\begin{align}
    \gamma _{t,small}^L = \frac{{{a_s}{P_B}{L_{BU,L}}(d_{0,j}){h_{0,j}^L}^2}}{{{I_L} + {I_N} + {\sigma ^2}}}.
\end{align}

We denote $d_{0,j}^{(i)}=d_{BR,j}^{(i)} d_{RU,0}^{(i)}$. If the typical UE $u_0 \in \Phi _U^R$, i.e. $d_{0,j}^{(i)} \le {( {{\tilde c}_{RL}})^{\frac{1}{\alpha _R}}}{{d_C}^{{\tilde \alpha }_{LR}}}$, the SINR for the SIC process at $u_0$ is given by
\begin{align}
    \gamma _{t \to c,small}^R = \frac{{{a_l}{P_B}L_{RIS}(d_{BR,j}^{(i)}, d_{RU,0}^{(i)}) h{{_{0,j}^{(i)}}^2}}}{{{a_s}{P_B}L_{RIS}(d_{BR,j}^{(i)}, d_{RU,0}^{(i)}) h{{_{0,j}^{(i)}}^2} + {I_R} + {I_L} + {I_N} + {\sigma ^2}}},
\end{align}
where ${I_R} = \sum\nolimits_{k \in \Phi _B^R\backslash j} {{P_B}L_{RIS}(d_{BR,k}^{(i)}, d_{RU,0}^{(i)})h{{_{0,k}^{(i)}}^2}} $ is the interference from BSs located at the same side of serving RIS $i$. We assume that only the serving RIS is pointed at the typical UE, so the links between the other RISs and the typical UE are blocked. The ${h_{0,k}^{(i)}}^2$ is the small scale fading power from the BS $k \in \Phi_B$, which also follows Nakagami-$m$ distribution with an integer parameter $m_R$.

Thus, the decoding SINR at the typical UE $u_0$ can be expressed as
\begin{align}
\gamma _{t,small}^R = \frac{{{a_s}{P_B}L_{RIS}(d_{BR,j}^{(i)}, d_{RU,0}^{(i)})h{{_{0,j}^{(i)}}^2}}}{{{I_R} + {I_L} + {I_N} + {\sigma ^2}}}.
\end{align}

\subsubsection{Large Path Loss Case}
When the typical UE has larger path loss, the SIC process occurs at the connected UE, while the signal transmitted to the connected UE is regarded as interference at the typical UE. In this case,  $a_t = a_l$ and $a_c = a_s$.

If the typical UE $u_0 \in \Phi _U^L$, i.e. ${d_{0,j}} > {d_C}$, the decoding SINR for $u_0$ is as follows
\begin{align}
    \gamma _{t,large}^L = \frac{{{a_l}{P_B}{L_{BU,L}}(d_{0,j}){h_{0,j}^L}^2}}{{{a_s}{P_B}{L_{BU,L}}(d_{0,j}){h_{0,j}^L}^2 + {I_L} + {I_N} + {\sigma ^2}}}.
\end{align}

On the other hand, If the typical UE $u_0 \in \Phi _U^R$, i.e. $d_{0,j}^{(i)} >  {( {{\tilde c}_{RL}})^{\frac{1}{\alpha _R}}}{{d_C}^{{\tilde \alpha }_{LR}}}$, the decoding SINR for $u_0$ can be expressed as
\begin{align}
    \gamma _{t,large}^R = \frac{{{a_l}{P_B}L_{RIS}(d_{BR,j}^{(i)}, d_{RU,0}^{(i)})h{{_{0,j}^{(i)}}^2}}}{{{a_s}{P_B}L_{RIS}(d_{BR,j}^{(i)}, d_{RU,0}^{(i)})h{{_{0,j}^{(i)}}^2} + {I_R} + {I_L} + {I_N} + {\sigma ^2}}}.
\end{align}

For the connected UE, since its location and the association choice are known, the expression of SINR is much simpler than that for typical UE. As the performance of the connencted UE has been investigated in \cite{1rismodel}, we only focus on the typical UE in this work.

\section{Path Loss and Association Analysis in RIS HetNets}
In this section, we first derive the expressions of some relevant distance distributions, based on which the UE association probability is investigated. Then, we provide the angle distributions for RIS-aided links, which are also fundamental for subsequent coverage performance analysis.

\subsection{Distance Distributions of the Nearest BS}
For the typical UE $u_0 \in \Phi _U$, BSs in the system can be divided into two independent sets: LoS BSs $\Phi _B^L$ and NLoS BSs $\Phi _B^N$. According to the ``Thinning Theorem'' \cite[Theorem 2.36]{7Stochastic}, the locations of LoS BSs and NLoS BSs obey inhomogeneous PPPs of density $\lambda _B^L(d_0) = {\lambda _B}{p_L}(d_0)$ and $\lambda _B^N(d_0) = {\lambda _B}{p_N}(d_0)$ at distance $d_0$, respectively. As the typical UE can associate to a LoS BS when at least one LoS transmission link exists, the following lemma first provides the probability that the set $\Phi _B^L$ is non-empty.
\begin{lemma}
The probability that the typical UE has at least one LoS BS is as follows
\begin{align}\label{eq: LoS BS exist}
    P_L = 1 - \exp \left( { - \frac{{2\pi {\lambda _B}}}{{{\beta ^2}}}} \right).
\end{align}
\end{lemma}
\begin{IEEEproof}
Based on \cite[Theorem 8]{8loslemma}, the probability $P_L$ can be expressed as ${P_L} =1-\mathop {\lim }\limits_{x \to \infty } \mathbb{P}({d_{0,j}} > x) =1 - \exp \left( { - 2\pi \int_0^\infty  {r{\lambda _B^L(r)}dr} } \right)$. By plugging \eqref{LoS probability}, \eqref{eq: LoS BS exist} is obtained.
\end{IEEEproof}

Thus, the probability that all LoS links are blocked for the typical UE is $P_{N}=1-P_L=\exp \left( { - \frac{{2\pi {\lambda _B}}}{{{\beta ^2}}}} \right)$. Then, the distribution of the distance to the nearest BS in $\Phi _B^L$ is calculated as below.

\begin{lemma}\label{lemma: PDF LoS}
For BS-UE links, the PDF of the distance between the typical UE and its nearest LoS BS $d_{0,j}$ can be given by
\begin{align}\label{eq: PDF LoS}
    {f_{{d_{0,j}}}}(x) = 2\pi {\lambda _B}x{p_L}(x)\exp \left( { - 2\pi {\lambda _B}\int_0^x {r{p_L}(r)dr} } \right).
\end{align}
\end{lemma}

\begin{IEEEproof}
 Using the feature of the inhomogeneous PPP $\Phi_B^L$, the cumulative distribution function (CDF) of $d_{0,j}$ is given by
\begin{align}\label{eq: CDF LoS}
    {F_{{d_{0,j}}}}\left( x \right) =  {1 - \exp \left( { - 2\pi {\lambda _B}\int_0^x {r{p_L}(r)dr} } \right)}.
\end{align}

The PDF of $d_{0,j}$ can be calculated by ${f_{{d_{0,j}}}}(x) = \frac{d}{{dx}}{F_{{d_{0,j}}}}(x)$. Thus, the PDF is obtained as \eqref{eq: PDF LoS}.
\end{IEEEproof}

Noticed that the locations of UEs, RISs, and BSs follow independent homogeneous PPPs, the PDF of the distance between a RIS and its nearest BS ${d_{BR,j}^{(i)}}$ as well as the PDF of the distance between the typical UE and its nearest RIS ${d_{RU,0}^{(i)}}$ can be given by
\begin{align}\label{PDF d_BR}
{f_{d_{BR}}}(x) = 2\pi {\lambda _B}x\exp \left( { - \pi {\lambda _B}{x^2}} \right),
\end{align}
\begin{align}\label{PDF d_RU}
    {f_{d_{RU}}}(x) = 2\pi {\lambda _R}x\exp \left( { - \pi {\lambda _R}{x^2}} \right).
\end{align}

\begin{lemma}\label{lemma: PDF d_BU_RIS}
For BS-RIS-UE links, the PDF of the distance between the typical UE and its nearest BS $d_{0,j}^{(i)}$ can be given by
\begin{align}\label{eq: PDF d_BU_RIS1}
    {f_{d_{0,j}^{(i)}}}(x) = 4{\pi ^2}x{\lambda _B}{\lambda _R}{K_0}\left( {2\pi x\sqrt {{\lambda _B}{\lambda _R}} } \right),
\end{align}
where ${K_0}( \cdot )$ is modified Bessel function of the second kind \cite[eq. (8.447.3)]{5}.
\end{lemma}
\begin{IEEEproof}
See Appendix A.
\end{IEEEproof}

\subsection{UE Association Probability}
As mentioned in \ref{model: UE Association}, the typical UE chooses the communication link with maximum average received power. The following Theorem provides the UE probability for each tier in this RIS HetNets.

\begin{theorem}\label{theorem: association probability}
The probability that the typical UE associates to a LoS BS, defined as $A_L = \mathbb{P}(u_0 \in \Phi _U^L)$, can be calculated as
\begin{align}\label{association probability AR1}
   {A_L} = 4{\pi ^2}{\lambda _B}{\lambda _R}\int_0^\infty  {x\left( {1 - \exp \left( { - 2\pi {\lambda _B}\int_0^{\varphi (x)} {r{p_L}(r)dr} } \right)} \right){K_0}\left( {2\pi x\sqrt {{\lambda _B}{\lambda _R}} } \right)dx},
\end{align}
where $\varphi (x) = {( {{{\tilde c}_{LR}}})^{\frac{1}{{{\alpha _L}}}}}{x^{{{\tilde \alpha }_{RL}}}}$. Thus, the probability that the typical UE connects to a RIS is $A_R = \mathbb{P}(u_0 \in \Phi _U^R) = 1-A_L$.
\end{theorem}
\begin{IEEEproof}
When the typical UE obtains larger received power from a LoS BS rather than a RIS, i.e.
$P_{0,j}>P_{0,j}^{(i)}$, the typical UE associates to a LoS BS through BS-UE link. Therefore,
\begin{align}\label{eq: AL proof}
    {A_L} &= \mathbb{P}\left( {{a_t}{P_B}{C_L}{d_{0,j}}^{ - {\alpha _L}} \ge {a_t}{P_B}{C_R}d{{_{BU,j}^{(i)}}^{ - {\alpha _R}}}} \right)\nonumber\\
    &= \mathbb{P}\left( {{d_{0,j}} \le {{\left(  {{{\tilde c}_{LR}}} \right)}^{\frac{1}{\alpha _L}}}d{{_{0,j}^{(i)}}^{{\tilde \alpha }_{RL}}}} \right)\nonumber\\
    &= \int_0^\infty  {\mathbb{P}\left( {{d_{0,j}} \le {{\left( {{{\tilde c}_{LR}}} \right)}^{\frac{1}{\alpha _L}}}{x^{{\tilde \alpha }_{RL}}}} \right){f_{d_{0,j}^{(i)}}}(x)dx}.
\end{align}

Based on \eqref{eq: CDF LoS} and \eqref{eq: PDF d_BU_RIS1}, the results in \eqref{association probability AR1} can be obtained.
\end{IEEEproof}

\begin{remark}\label{remark: bounds for association prob}
When $x<\infty$, $\varphi (x)< \infty$ always holds. The upper bound of $A_L$ can be calculated as
\begin{align}
    {A_L} &< 4{\pi ^2}{\lambda _B}{\lambda _R}\int_0^\infty  {x\left( {1 - \exp \left( { - 2\pi {\lambda _B}\int_0^{\infty} {r{p_L}(r)dr} } \right)} \right){K_0}\left( {2\pi x\sqrt {{\lambda _B}{\lambda _R}} } \right)dx}\nonumber
    \\ & = 4{\pi ^2}{\lambda _B}{\lambda _R}\int_0^\infty  {xP_L{K_0}\left( {2\pi x\sqrt {{\lambda _B}{\lambda _R}} } \right)dx}\nonumber\\&=P_L.
\end{align}
Thus, the lower bound of $A_R$ is $P_{N}$. It can be explained that the UEs associate to BSs through BS-RIS-UE links when all LoS BS-UE links are blocked. According to \eqref{eq: LoS BS exist}, in the sparse BS or dense blockage environment, the probability that a UE is blocked is high, so most UEs aquire signal from RISs for larger received power.
\end{remark}

\begin{corollary}
For the special case that $p_L(x)=1$ and $\alpha _L = 2\alpha _R$, the UE association probability for $u_0 \in \Phi _U^L$ can be expressed in closed form as
\begin{align}\label{eq: corollary 1}
    {A_L} = 1 - \frac{{4{{\tilde \lambda }_{RB}}}}{{{c_{LR}}^4 - 4{{\tilde \lambda }_{RB}}}}\left( {\frac{{{c_{LR}}^2}}{{\sqrt {{c_{LR}}^4 - 4{{\tilde \lambda }_{RB}}} }}\ln \left( {\frac{{{c_{LR}}^2}}{{2\sqrt {{{\tilde \lambda }_{RB}}} }} + \sqrt {\frac{{{c_{LR}}^4}}{{4{{\tilde \lambda }_{RB}}}} - 1} } \right) - 1} \right),
\end{align}
where ${c_{LR}} = {\left( {{{\tilde c}_{LR}}} \right)^{\frac{1}{{{\alpha _L}}}}}$ and ${{\tilde \lambda }_{RB}} = \lambda _R / \lambda _B$.
\end{corollary}
\begin{IEEEproof}
In this case, $\varphi (x) = {\left( {{{\tilde c}_{LR}}} \right)^{\frac{1}{{{\alpha _L}}}}}{x^{\frac{1}{2}}}$ and \eqref{association probability AR1} can be rewritten as
\begin{align}
    {A_L} = 4{\pi ^2}{\lambda _B}{\lambda _R}\int_0^\infty  {x\left( {1 - \exp \left( { - \pi {\lambda _B}{{({{\tilde c}_{LR}})}^{2/{\alpha _L}}}{x}} \right)} \right){K_0}\left( {2\pi x\sqrt {{\lambda _B}{\lambda _R}} } \right)dx} .
\end{align}

By using eq. (6.561.16) and eq. (6.624.1) in \cite{5}, \eqref{eq: corollary 1} is obtained.
\end{IEEEproof}

\begin{corollary}
For the special case that $p_L(x)=1$ and $\alpha _L = \alpha _R$, the UE association probability for $u_0 \in \Phi _U^L$ can be expressed in closed form as
\begin{align}\label{eq: corollary 2}
     {A_L} = 1 - \frac{{\sqrt {\pi {\lambda _R}} }}{{{c_{LR}}}}\exp (\frac{{\pi {\lambda _R}}}{{2{c_{LR}}^2}}){W_{ - \frac{1}{2},0}}(\frac{{\pi {\lambda _R}}}{{{c_{LR}}^2}}),
\end{align}
where ${W_{ \cdot , \cdot }}( \cdot )$ is the Whittaker function \cite[eq. (9.220.2)]{5}.
\end{corollary}
\begin{IEEEproof}
In this case, $\varphi (x) = {\left( {{{\tilde c}_{LR}}} \right)^{\frac{1}{{{\alpha _L}}}}}{x}$. Using eq. (6.561.16) and eq. (6.631.3) in \cite{5}, this corollary is proved.
\end{IEEEproof}

\begin{remark}
In these special cases, the value of $P_L$ is related to ${\tilde \lambda }_{RB}$ or $\lambda _R$ rather than the exact density of BSs $\lambda _B$. Thus, in these non-blockage environments, the received power at UEs can also be improved by deploying denser RISs.
\end{remark}

\subsection{Angle Distribution}
As shown in \eqref{model: RIS path loss}, for the typical UE, the path loss expression of BS-RIS-UE link relates to the angle of arrival $\theta _{BR,0}$ and the angle of departure $\theta _{RU,0}$. Since the reflective surfaces are regarded as RISs rather than mirrors, $\theta _{BR,0}$ and $\theta _{RU,0}$ can be unequal.

According to \cite[Remark 1]{1rismodel}, the angle $\theta = \theta _{BR,0}+ \theta _{RU,0}$ is uniformly distributed in $[0,\pi]$. We denote $\varepsilon_0 \in (0,1)$, so the angle of arrival and the angle of departure can be expressed as $\theta _{BR,0} = \varepsilon_0 \theta$ and $\theta _{RU,0} = (1-\varepsilon_0) \theta$, respectively. The PDFs of the angle of arrival and departure are as follows
\begin{align}
    {f_{{\theta _{BR,0}}}}(x)&= \frac{1}{{\pi {\varepsilon _0}}},\quad x \in (0,\frac{\pi }{2}), \\
    {f_{{\theta _{RU,0}}}}(x) &= \frac{1}{{\pi (1 - {\varepsilon _0})}},\quad x \in (0,\frac{\pi }{2}).
\end{align}

As the angles of arrival and departure obey uniform distributions, $C_R$ can be approximated as the average value
\begin{align}
    {C_R} \approx \mathbb{E}\left[ {{C_R}} \right] = \frac{{{L^2}}}{{16{\pi ^3}}}\left( {\pi  + \frac{{\sin (2{\varepsilon _0}\pi )}}{{4{\varepsilon _0} - 12{\varepsilon _0}^2 + 8{\varepsilon _0}^3}}} \right).
\end{align}

The intercept of the path loss for BS-RIS-UE links has a positive correlation with the length of RISs. Thus, we can use larger RISs to enhance the channel conditions.

\section{Coverage Probability Analysis}
The coverage probability is generally defined as the probability that the typical UE can successful transmit signals with a targeted SINR $\tau _t$ or a targeted data rate $\rho _t$. In this section, we provide both the SINR coverage probability and rate coverage probability of the typical UE $u_0 \in \Phi _U$.

\subsection{Laplace Transform of Interference}
Before analyzing the coverage performance of this system, three kinds of Laplace transforms of interference are derived first. Only active BSs become the interfering BSs. Let $I _{total} = I_L + I_N +I_R$ denote the total interference to the typical UE, where $I_L$, $I_N$, and $I_R$ are the interference from LoS BSs, NLoS BSs, and RISs, respectively. The Laplace transform of $I _{total}$ is ${\cal L}_{I _{total}}(s) = {\cal L}_{I _{L}}(s){\cal L}_{I _{NL}}(s){\cal L}_{I _{R}}(s)$.

\subsubsection{Interference from LoS BSs}
For both conditions that the typical UE associates to a LoS BS and the typical UE associates to a RIS, the Laplace transform of the interference from LoS BSs can be expressed as
\begin{align}
   {{\cal L}_{{I_L}}}(s) = \mathbb{E}\left[ {\exp \left( { - s\sum\limits_{k \in \Phi _B^L\backslash j} {{P_B}{L_{BU,L}}(d_{0,k})h{{_{0,k}^L}^2}} } \right)} \right].
\end{align}

\begin{lemma}\label{Lemma: Interference LoS BSs}
The Laplace transform of the interference from LoS BSs is derived as
\begin{align}
   {{\cal L}_{{I_L}}}(s) = \exp \left( { - 2\pi {\tilde \lambda _B}\int_{d_{0,min}}^\infty  {\left( {1 - {{\left( {1 + \frac{{s{P_B}{L_{BU,L}}(x)}}{m_L}} \right)}^{ - m_L}}} \right)x{p_L}(x)} {\rm{d}}x} \right),
\end{align}
where ${d_{0,min }} = \min \left\{ {{d_{0,j}},\varphi\left({d_{0,j}^{(i)}}\right)}\right\} $ is the minimum distance between the typical UE and interfering BSs.
\end{lemma}
\begin{IEEEproof}
See Appendix B.
\end{IEEEproof}

\subsubsection{Interference from NLoS BSs}
Similarly, the Laplace transform of the interference from NLoS BSs can be expressed as
\begin{align}
   {{\cal L}_{{I_N}}}(s) = \mathbb{E}\left[ {\exp \left( { -s \sum\limits_{k \in \Phi _B^N\backslash j} {{P_B}{L_{BU,N}}(d_{0,k})h{{_{0,k}^{N}}^2}} } \right)} \right].
\end{align}

\begin{lemma}\label{Lemma: Interference NLoS BSs}
The Laplace transform of the interference from NLoS BSs is derived as
\begin{align}
   {{\cal L} _{{I_N}}}(s) = \exp \left( { - 2\pi {\tilde \lambda _B}\int_0^\infty  {\left( {1 - {{\left( {1 + \frac{{s{P_B}{L_{BU,N}}(x)}}{{{m_N}}}} \right)}^{ - {m_N}}}} \right)x{p_N}(x)} {\rm{d}}x} \right).
\end{align}
\end{lemma}
\begin{IEEEproof}
Since the typical UE can not associate to a NLoS BS directly, the minimum distance between the typical UE and interfering NLoS BSs is $0$. The derivation procedure is the same as the proof in \textbf{Lemma~\ref{Lemma: Interference LoS BSs}}.
\end{IEEEproof}

\subsubsection{Interference from RISs}
If the typical UE associates to a BS through BS-RIS-UE link, it receives signal as well as interference reflected from the serving RIS. Under this condition, the Laplace transform of the interference from the RIS can be expressed as
\begin{align}
    {{\cal L} _{{I_R}}}(s) = \mathbb{E}\left[ {\exp \left( { -s \sum\limits_{k \in \Phi _B^R\backslash j} {{P_B}L_{RIS}(d_{BR,k}^{(i)}, d_{RU,0}^{(i)})h{{_{0,k}^R}^2}} } \right)} \right].
\end{align}

\begin{lemma}\label{Lemma: Interference RISs}
The Laplace transform of the interference from the assistant RIS is derived as
\begin{align}
    {{\cal L} _{{I_R}}}(s) = \exp \left( { - {\delta _1}\left( {_2{F_1}\left( { {m_R},- \frac{2}{{{\alpha _R}}};1 - \frac{2}{{{\alpha _R}}}; - s{\delta _2}} \right) - 1} \right)} \right),
\end{align}
where ${\delta _1 = \frac{{\pi {\tilde \lambda _B} d {{_{BR,j}^{(i)}}^2}}}{2}}$, $\delta _2 = {\frac{{{P_B}L_{RIS}(d_{BR,j}^{(i)}, d_{RU,0}^{(i)})}}{{{m_R}}}}$, and $_2{F_1}\left( { \cdot , \cdot ; \cdot ; \cdot } \right)$ is the Gauss hypergeometric function.
\end{lemma}
\begin{IEEEproof}
See Appendix C.
\end{IEEEproof}

\subsection{SINR Coverage Probability}
Based on the distributions of the relevant distance $d_{0,j}$, $d_{BR,j}^{(i)}$, and $d_{RU,0}^{(i)}$, the association probabilities $A_L$, and the Laplace transforms of interference $I_L$, $I_N$, and $I_R$, we proceed
to derive the analytical expression of the SINR coverage probability. According to the path losses, two cases are considered in the following.

\subsubsection{Small Path Loss Case}
In this case, the typical UE decodes its own message after a successful SIC process, and the SINR coverage probability is expressed as
\begin{align}
    {P_{{\mathop{cov}} ,small}} = \mathbb{P}\left( {{\gamma _{t \to c,small}} > {\tau _c},{\gamma _{t,small}} > {\tau _t}} \right),
\end{align}
where $\tau _c$ and $\tau _t$ are the targeted SINRs of the connected UE and the typical UE, respectively.

If the typical UE $u_0 \in \Phi _U^L$ and
${{a_l} - {\tau _c}{a_s}} > 0$, the SINR coverage probability can be rewritten as
\begin{align}
    P_{cov ,small}^L({d_{0,j}}) = \mathbb{P}\left( {{h_{0,j}}^2 > \frac{{{\tau ^ * }\left( {{I_L} + {I_N} + {\sigma ^2}} \right)}}{{{P_B}{L_{BU,L}}(d_{0,j})}}} \right),
\end{align}
where ${\tau ^ * } = \max \left( {\frac{{{\tau _c}}}{{{a_l} - {\tau _c}{a_s}}},\frac{{{\tau _t}}}{{{a_s}}}} \right)$.

\begin{lemma}\label{lemma: small LoS CP}
If ${{a_l} - {\tau _c}{a_s}} > 0$ holds, the approximated SINR coverage probability of the typical UE $u_0 \in \Phi _U^L$ for the small path loss case is derived as
\begin{align}\label{eq: small LoS CP}
    P_{{\mathop{\rm cov}} ,small}^L({d_{0,j}})
    \approx \sum\limits_{n = 1}^{{m_L}} {{{( - 1)}^{n + 1}}} {\binom{m_L}{n}} {{\cal L}_{{I_L}}}({s_L}){{\cal L}_{{I_N}}}({s_L}){e^{ - {s_L}{\sigma ^2}}},
\end{align}
where ${s_L} = \frac{{n{\eta _L}{\tau ^ * }}}{{{P_B}{L_{BU,L}}(d_{0,j})}}$ and $\eta _L= m_L{(m_L!)^{ - \frac{1}{m_L}}}$. Otherwise, $P_{{\mathop{\rm cov}} ,small}^L({d_{0,j}}) =0$.
\end{lemma}

\begin{IEEEproof}
According to \cite{5Gammalemma}, the normalized Gamma variable $h^2$ with parameter $m$ has a tight lower bound $\mathbb{P}({h^2} < x) > {\left( {1 - {e^{ - \eta x}}} \right)^m}$, where $\eta  = m{(m!)^{ - \frac{1}{m}}}$. Utilizing binomial expansions, \eqref{eq: small LoS CP} is obtained.
\end{IEEEproof}

If the typical UE $u_0 \in \Phi _U^R$ and
${{a_l} - {\tau _c}{a_s}} > 0$, the SINR coverage probability can be rewritten as
\begin{align}
    P_{cov,small}^R(d_{BR,j}^{(i)},d_{RU,0}^{(i)}){\rm{ }} = \mathbb{P}\left( {h{{_{0,j}^{(i)}}^2} > \frac{{{\tau ^*}\left( {{I_R} + {I_L} + {I_N} + {\sigma ^2}} \right)}}{{{P_B}L_{RIS}(d_{BR,j}^{(i)}, d_{RU,0}^{(i)})}}} \right).
\end{align}
%where $\zeta ({d_C}) = {{\tilde C}_{RL}}^{1/{\alpha _R}}{d_C}^{{{\tilde \alpha }_{LR}}}$.

\begin{lemma}\label{lemma: small RIS CP}
If ${{a_l} - {\tau _c}{a_s}} > 0$ holds, the approximated SINR coverage probability of the typical UE $u_0 \in \Phi _U^R$ for the small path loss case is derived as
\begin{align}
     P_{cov,small}^R(d_{BR,j}^{(i)},d_{RU,0}^{(i)})
     \approx \sum\limits_{n = 1}^{{m_R}} {{{( - 1)}^{n + 1}}} {\binom{m_R}{n}}{{\cal L}_{{I_R}}}({s_R}){{\cal L}_{{I_L}}}({s_R}){{\cal L}_{{I_N}}}({s_R}){e^{ - {s_R}{\sigma ^2}}},
\end{align}
where ${s_R} = \frac{{n{\eta _R}{\tau ^ * }}}{{{P_B}L_{RIS}(d_{BR,j}^{(i)}, d_{RU,0}^{(i)})}}$. Otherwise, $P_{cov,small}^R(d_{BR,j}^{(i)},d_{RU,0}^{(i)}) =0$.
\end{lemma}

\begin{IEEEproof}
The proof is similar to the proof in \textbf{Lemma~\ref{lemma: small LoS CP}}.
\end{IEEEproof}

\subsubsection{Large Path Loss Case}
In this case, the typical UE decodes its own message by treating the paired connected UE as noise, so the SINR coverage probability is
\begin{align}
    {P_{{\mathop{cov}} ,large}} = \mathbb{P}\left( {{\gamma _{t,large}} > {\tau _t}} \right).
\end{align}

Similar to the small path loss case, the conditional SINR coverage probabilities of the typical user for the large path loss case are given in the following propositions.
\begin{proposition}\label{proposition: large LoS CP}
If ${{a_l} - {\tau _t}{a_s}} > 0$ holds, the approximated  SINR coverage probability of the typical UE $u_0 \in \Phi _U^L$ for the large path loss case is derived as
\begin{align}
    P_{cov ,large}^L({d_{0,j}}) \approx \sum\limits_{n = 1}^{{m_L}} {{{( - 1)}^{n + 1}}} {\binom{m_L}{n}}{{\cal L}_{{I_L}}}(s_L^l){{\cal L}_{{I_N}}}(s_L^l){e^{ - s_L^l{\sigma ^2}}},
\end{align}
where $s_L^l = \frac{{n{\eta _L}\tau _t^l}}{{{P_B}{L_{BU,L}}(d_{0,j})}}$ and $\tau _t^l = \frac{{{\tau _t}}}{{{a_l} - {\tau _t}{a_s}}}$. Otherwise, $P_{cov ,large}^L({d_{0,j}})=0$.
\end{proposition}

\begin{proposition}\label{proposition: large RIS CP}
If ${{a_l} - {\tau _t}{a_s}} > 0$ holds, the approximated  SINR coverage probability of the typical UE $u_0 \in \Phi _U^R$ for the large path loss case is derived as
\begin{align}
    P_{cov,large}^R(d_{BR,j}^{(i)},d_{RU,0}^{(i)})
     \approx \sum\limits_{n = 1}^{{m_R}} {{{( - 1)}^{n + 1}}} {\binom{m_R}{n}}{{\cal L}_{{I_R}}}(s_R^l){{\cal L}_{{I_L}}}(s_R^l){{\cal L}_{{I_N}}}(s_R^l){e^{ - s_R^l{\sigma ^2}}},
\end{align}
where $s_R^l = \frac{{n{\eta _R}\tau _t^l}}{{{P_B}L_{RIS}(d_{BR,j}^{(i)}, d_{RU,0}^{(i)})}}$. Otherwise, $P_{cov,large}^R(d_{BR,j}^{(i)},d_{RU,0}^{(i)}) =0$.
\end{proposition}

\begin{remark}
	Comparing to the case that the typical UE associates to a LoS BS directly, the typical UE suffers extra interference $I_R$ when associating to a RIS. Thus, when the pass losses for the two kinds of links are similar in value, the typical UE obtains higher SINR from the BS-UE link than the BS-RIS-UE link. In practical application scenarios, we can set a bias factor in the UE association scheme to improve the system performance.
\end{remark}

Now that we have developed expressions of the conditional SINR coverage probability in \textbf{Lemma~\ref{lemma: small LoS CP}-\ref{lemma: small RIS CP}} and \textbf{Proposition~\ref{proposition: large LoS CP}-\ref{proposition: large RIS CP}}, the SINR coverage probability of the typical UE can be calculated in the following theorem.

\begin{theorem}\label{theorem: L/R UE cp}
For the typical UE in this NOMA RIS HetNets, the SINR coverage probability is expressed as
\begin{align}\label{eq: cp}
    {P_{cov }}({\tau _c},{\tau _t}) = P_{cov }^L({\tau _c},{\tau _t}) + P_{cov }^R({\tau _c},{\tau _t}).
\end{align}
$P_{cov }^L({\tau _c},{\tau _t})$ and $P_{cov }^R({\tau _c},{\tau _t})$ are the SINR coverage probabilities when the typical UE is associated with a LoS BS and RIS, respectively, and are given by
\begin{align}
    &P_{cov }^L({\tau _c},{\tau _t}) = \int_0^{{d_C}} {P_{cov,small}^L(x){f_{{d_{0,j}}}}(x)\int_{{\varphi ^{ - 1}}(x)}^\infty  {{f_{d_{0,j}^{(i)}}}(y)dy} } dx \nonumber \\
    &+ \int_{{d_C}}^\infty  {P_{cov,large}^L(x)} {f_{{d_{0,j}}}}(x)\int_{{\varphi ^{ - 1}}(x)}^\infty  {{f_{d_{0,j}^{(i)}}}(y)dy} dx,
\end{align}
\begin{align}\label{eq: cp for RIS}
    &P_{cov}^R({\tau _c},{\tau _t}) = \int_0^\infty  {\int_0^{\varphi^{-1} ({d_C})/{x_1}} {P_{cov,small}^R\left( {{x_1},{x_2}} \right){{\bar F}_{d_{0,j}}}\left( {\varphi \left( {{x_1}{x_2}} \right)} \right){f_{{d_{BR}}}}\left( {{x_2}} \right)d{x_2}} } {f_{{d_{RU}}}}\left( {{x_1}} \right)d{x_1}\nonumber \\
    &  + \int_0^\infty  {\int_{\varphi^{-1} ({d_C})/{x_1}}^\infty  {P_{cov,large}^R\left( {{x_1},{x_2}} \right){{\bar F}_{d_{0,j}}}\left( {\varphi \left( {{x_1}{x_2}} \right)} \right){f_{{d_{BR}}}}\left( {{x_2}} \right)d{x_2}} } {f_{{d_{RU}}}}\left( {{x_1}} \right)d{x_1},
\end{align}
where $\varphi^{-1} (x) = {\left( {{{\tilde c}_{RL}}} \right)^{\frac{1}{{{\alpha _R}}}}}{x^{{{\tilde \alpha }_{LR}}}}$, and ${\bar F}_{d_{0,j}}(x)=1-F_{d_{0,j}}(x)$.
\end{theorem}

\begin{IEEEproof}
See Appendix D.
\end{IEEEproof}

Based on the analytical expression of the SINR coverage probability provided in \textbf{Theorem~\ref{theorem: L/R UE cp}}, we consider a special case in the RIS HetNets.

\begin{corollary}\label{corollary: Large length L}
	Conditioned on the half-length of RISs $L \to \infty$, the asymptotic SINR coverage probability of the typical UE is
	\begin{align}
	&P_{cov}\left( {{\tau _c},{\tau _t}} \right) \approx \sum\limits_{n = 1}^{{m_R}} {{{( - 1)}^{n + 1}}} {\binom{m_R}{n}}\frac{{{\lambda _B}}}{{\frac{{{{\tilde \lambda }_B}}}{2}{Q_R}\left( {{\eta _R},{\tau ^*}} \right) + {\lambda _B}}},
	\end{align}
	where ${{Q_R}\left( {\eta ,\tau } \right)}= {}_2{F_1}\left( {{m_R}, - \frac{2}{{{\alpha _R}}};1 - \frac{2}{{{\alpha _R}}}; - \frac{{n\eta \tau }}{{{m_R}}}} \right) - 1$.
\end{corollary}
\begin{IEEEproof}
See Appendix E.
\end{IEEEproof}

\begin{remark}\label{Remark: RIS enlargement}
When the length of RISs is sufficiently large, the SINR coverage becomes a constant related to the density of BSs $\lambda_B$ and the density of active BSs ${\tilde \lambda }_B$. Moreover, if $\lambda_U \gg \lambda_B$, ${\tilde \lambda }_B \to \lambda_B$ and the SINR coverage $P_{cov}\left( {{\tau _c},{\tau _t}} \right) <1$. If $\lambda_U \ll \lambda_B$, the density of active BSs ${\tilde \lambda }_B \to \lambda_U$ and the part $\frac{{{{\tilde \lambda }_B}}}{2}{Q_R}\left( {{\eta _R},{\tau ^*}} \right)$ is negligible. In this case, $P_{cov}\left( {{\tau _c},{\tau _t}} \right) \to 1$. Thus, the enlargement of RISs may not achieve the maximum achievable performance.
\end{remark}

\subsection{Rate Distribution}
In this subsection, we evaluate the rate coverage probability of the typical UE. Noticed that the targeted data rate of the connected UE is $\rho _c = B_W \log _2(1+\gamma_c)$ and that of the typical UE is $\rho _t = B_W \log _2(1+\gamma_t)$, where $B_W$ is the available bandwidth, the expression of rate coverage probability can be derived from the SINR coverage probability.

Let $N_B$ denote the number of UEs that associate to a BS. According to \cite[Lemma 3]{6load}, the probability mass function (PMF) of the load at a BS is given by
\begin{align}\label{eq: PMF}
    {f_B}(n) = \mathbb{P}({N_B} = n) = \frac{{{{3.5}^{3.5}}\Gamma (n + 3.5)}}{{\Gamma (3.5)\Gamma (n)}}{\left( {\frac{{{\lambda _U}}}{{{\lambda _B}}}} \right)^{n - 1}}{\left( {3.5 + \frac{{{\lambda _U}}}{{{\lambda _B}}}} \right)^{ - (n + 3.5)}},
\end{align}
where $ n \ge 1 $ and $\Gamma (\cdot)$ is the gamma function.

\begin{theorem}\label{theorem: exact rate}
For the typical UE in the NOMA RIS HetNets, the rate coverage probability is given by
\begin{align}\label{eq: exact rate cv}
    {R_{cov}}({\rho _c},{\rho _t}) = \sum\limits_{n \ge 1} {{f_B}(n)} \left( {P_{cov}^L\left( {\phi (n{\rho _c}),\phi (n{\rho _t})} \right) + P_{cov}^R\left( {\phi (n{\rho _c}),\phi (n{\rho _t})} \right)} \right),
\end{align}
where $\phi (x) = {2^{x/W}} - 1$.
\end{theorem}
\begin{IEEEproof}
The probability that the data rate requirement of the typical UE $u_0 \in \Phi _U$ is met is expressed as
\begin{align}
    {R_{cov }}({\tau _c},{\tau _t})& = \sum\limits_{T = \{ L,R\} } {\mathbb{P}{\left( {\frac{W}{{{N_B}}}{{\log }_2}\left( {1 + \gamma _t^T} \right) > {\rho _t} \cap \frac{W}{{{N_B}}}{{\log }_2}\left( {1 + \gamma _{t \to c,small}^T} \right) > {\rho _c}} ,{{u_0} \in \Phi _U^T}\right)}}\nonumber\\
    &= \sum\limits_{T = \{ L,R\} } {\mathbb{P}\left( {\gamma _t^T > \phi \left( {{\rho _t}{N_B}} \right) \cap \gamma _{t \to c,small}^T > \phi \left( {{\rho _c}{N_B}} \right)},{{u_0} \in \Phi _U^T} \right)} \nonumber\\
    &= \sum\limits_{T = \{ L,R\} } {{\mathbb{E}_{{N_B}}}\left[ {P_{cov}^T\left( {\phi \left( {{\rho _c}{N_B}} \right),\phi \left( {{\rho _t}{N_B}} \right)} \right)} \right]},
\end{align}
where $\phi (x) = {2^{x/W}} - 1$. Using the result in \eqref{eq: PMF}, the average value is
\begin{align}
  {\mathbb{E}_{{N_B}}}\left[ {P_{cov}^T\left( {\phi \left( {{\rho _c}{N_B}} \right),\phi \left( {{\rho _t}{N_B}} \right)} \right)} \right] = \sum\limits_{n \ge 1} {{f_B}(n)} P_{cov}^T\left( {\phi \left( {n{\rho _c}} \right),\phi \left( {n{\rho _t}} \right)} \right).
\end{align}

The proof is completed.
\end{IEEEproof}

As shown in \eqref{eq: exact rate cv}, the expression of the rate coverage probability includes an infinite summation. For tractability of the analysis, we evaluate the rate coverage by introducing the mean load in the following corollary.
\begin{corollary}\label{corollary: approximate rate}
The rate coverage probability with the mean load approximation is given by
\begin{align}
    {{\bar R}_{cov}}({\rho _c},{\rho _t}) = P_{cov}^L\left( {\phi ({{\bar N}_B}{\rho _c}),\phi ({{\bar N}_B}{\rho _t})} \right) + P_{cov}^R\left( {\phi ({{\bar N}_B}{\rho _c}),\phi ({{\bar N}_B}{\rho _t})} \right).
\end{align}
\end{corollary}
\begin{IEEEproof}
The first moment of load $N_B$ is ${{\bar N}_B} =\mathbb{E}\left[ N_B\right]= 1 + \frac{{1.28{\lambda _U}}}{{{\lambda _B}}}$, \cite{6load}. Further, using the approximation ${\mathbb{E}_{{N_B}}}\left[ {P_{cov}^T\left( {\phi \left( {{\rho _c}{N_B}} \right),\phi \left( {{\rho _t}{N_B}} \right)} \right)} \right] \approx P_{cov}^T\left( {\phi \left( {{\rho _c}\mathbb{E}\left[ {{N_B}} \right]} \right),\phi \left( {{\rho _t}\mathbb{E}\left[ {{N_B}} \right]} \right)} \right)$, this corollary is proved.
\end{IEEEproof}

\begin{remark}
With the growth of $\lambda _U$, the load of each BS increases and the available bandwidth for each UE decreases. Hence, the rate coverage probability decreases.
\end{remark}

For the rate coverage, we also consider the special case $L \to \infty$ in the RIS HetNets.
\begin{corollary}\label{corollary: Large length L rate}
	Conditioned on the half-length of RISs $L \to \infty$, the asymptotic rate coverage probability of the typical UE is
	\begin{align}
	&R_{cov}\left( {{\tau _c},{\tau _t}} \right) \approx \sum\limits_{n = 1}^{{m_R}} {{{( - 1)}^{n + 1}}} {\binom{m_R}{n}}\frac{{{\lambda _B}}}{{\frac{{{{\tilde \lambda }_B}}}{2}{Q_R}\left( {{\eta _R},{\rho ^*}} \right) + {\lambda _B}}},
	\end{align}
	where $\rho ^* = \max \left( {\frac{{\phi ({{\bar N}_B}{\rho _c})}}{{{a_l} - \phi ({{\bar N}_B}{\rho _c}){a_s}}},\frac{{\phi ({{\bar N}_B}{\rho _t})}}{{{a_s}}}} \right)$.
\end{corollary}
\begin{IEEEproof}
	The proof is similar to \textbf{Corollary~\ref{corollary: Large length L}} and is hence skipped.
\end{IEEEproof}

\begin{remark}\label{Remark: RIS enlargement rate}
	Considering the rate coverage probability, the conclusion of RIS enlargement is the same as \textbf{Remark~\ref{Remark: RIS enlargement}}.
\end{remark}

\section{Numerical Results}
In this section, numerical results are presented to validate the UE association and coverage performance of NOMA enhanced RIS HetNets, and then some interesting insights are provided. We assume the network is operated at 28 GHz. The noise power is $\sigma^2 = -170+10\log_{10}W+N_f$. The distance bentween connected UEs and their BSs is fixed at $d_C=50$ m.  Table \ref{table: parameters} sunmmarizes the simulation parameters used in this section.
\begin{table}
\centering
\caption{Table of Parameters}\label{table: parameters}	
\begin{tabular}{m{3.2cm}<{\centering} |m{4cm}<{\centering}}
	\hline \hline
	\textbf{Bandwidth} & $W=100$ MHz \\ \hline
	\textbf{Path loss exponent} & $\alpha_L=2$, $\alpha_R=2.8$, $\alpha_{N}=4$ \\ \hline
	\textbf{Nakagami coefficient} & $m_L=m_R=4$, $m_N=1$ \\ \hline
	\textbf{Power allocation factor} & $a_s = 0.3$, $a_l = 0.7$ \\ \hline
	\textbf{Half-length of RIS} & $L=1$ m \\ \hline
	\textbf{Density of UE} & $\lambda_U = 100$ $\rm{km^{-2}}$ \\ \hline
	\textbf{Blockage parameter} & $\beta = 1/141.4$ \\ \hline
	\textbf{The noise figure} & $N_f=10$ dB \\ \hline
	\textbf{SINR threshold} & $\tau_t=\tau_c=-20$ dB\\ \hline
	\textbf{Rate threshold} & $\rho_t=\rho_c=1$ Mbps\\ \hline \hline
\end{tabular}
\end{table}
\vspace{-0.3 cm}
\subsection{UE Association Probability}
\begin{figure} [t!]
	\centering
	\includegraphics[width = 3.7in] {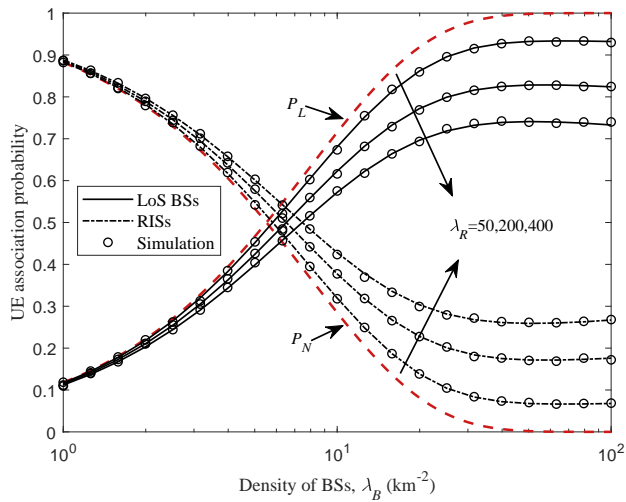}
	\caption{User association probability versus density of BSs with $\lambda_R=[50,200,400]$ km$^{-2}$: a verification of \textbf{Theorem~\ref{theorem: association probability}}.}
	\label{figure_association}
\end{figure}

Fig. \ref{figure_association} shows the effect of the density of BSs $\lambda_B$ and the effect of RISs $\lambda_R$ on UE association probability. The analytical curves presenting LoS BSs and RISs are from \textbf{Theorem~\ref{theorem: association probability}}, and the bounds $P_L$ and $P_{N}$ demonstrated in \textbf{Remark~\ref{remark: bounds for association prob}} are validated. One can observe that increasing the density of RISs encourages more UEs to associate to BSs through BS-RIS-UE links. This is because the RISs are closer to UEs and the reflective signal from RISs suffers smaller path loss than the signal from direct BS-UE links. Another observation is that with the increase of the BS density, the probability that the typical UE directly associates to LoS BSs increases to a maximum first and then keeps steadily. This can be explained that: 1) in sparse scenarios, deploying denser BSs can increase the probability that the BS-UE link exists, and this kind of link brings high received power;  2) in dense scenarios, increasing the density of BSs shortens the distance of both BS-UE links and BS-RIS-UE links, hence the received power from both tiers can be enhanced.
\begin{figure} [t!]
	\centering
	\includegraphics[width = 3.7in] {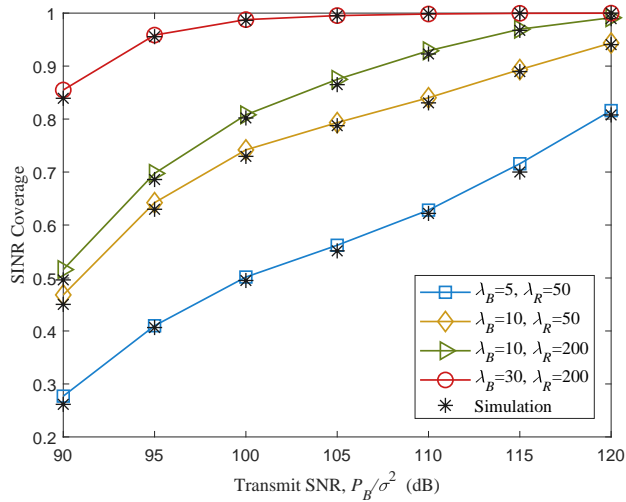}
	\caption{SINR coverage probability versus transmit SNR with various density of BSs and RISs: a verification of \textbf{Theorem~\ref{theorem: L/R UE cp}}.
	}
	\label{figure_Monte_carlo}
\end{figure}
\vspace{-0.2 cm}
\subsection{SINR Coverage}
Verification of the analytical expression of SINR coverage probability (\textbf{Theorem~\ref{theorem: L/R UE cp}}) is illustrated in Fig. \ref{figure_Monte_carlo} by sweeping over a range of transmit SNR. The SINR coverage is shown for different densities of BSs and RISs. It is observed that varying the density of BSs brings significant change in SINR coverage performance.

\begin{figure} [t!]
	\centering
	\includegraphics[width = 3.7in] {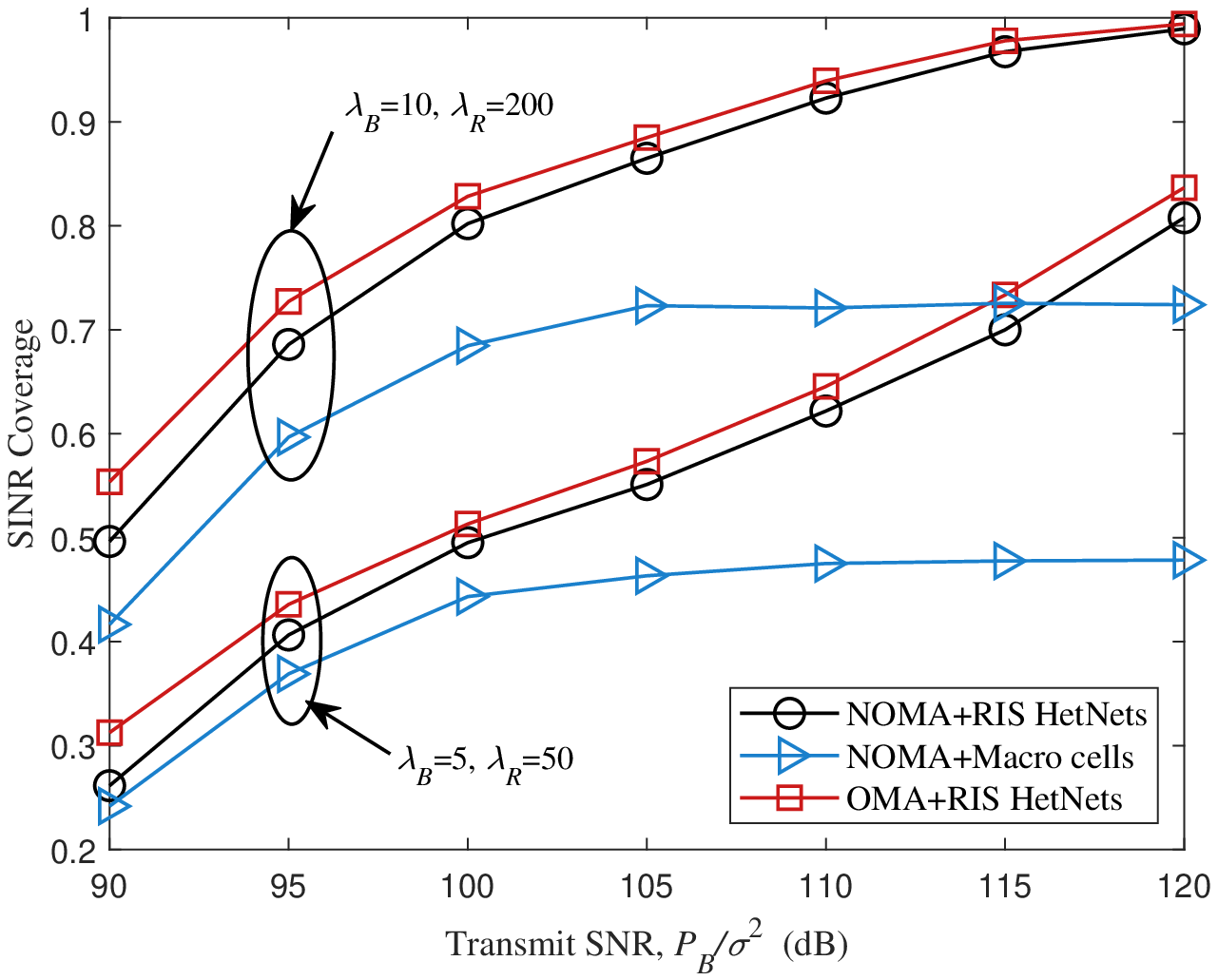}
	\caption{SINR coverage probability versus transmit SNR: a comparison among NOMA HetNets, OMA RIS HetNets and NOMA Macro cells.
	}
	\label{figure_SINR_comparison}
\end{figure}
\begin{figure} [t!]
	\centering
	\includegraphics[width = 3.7in] {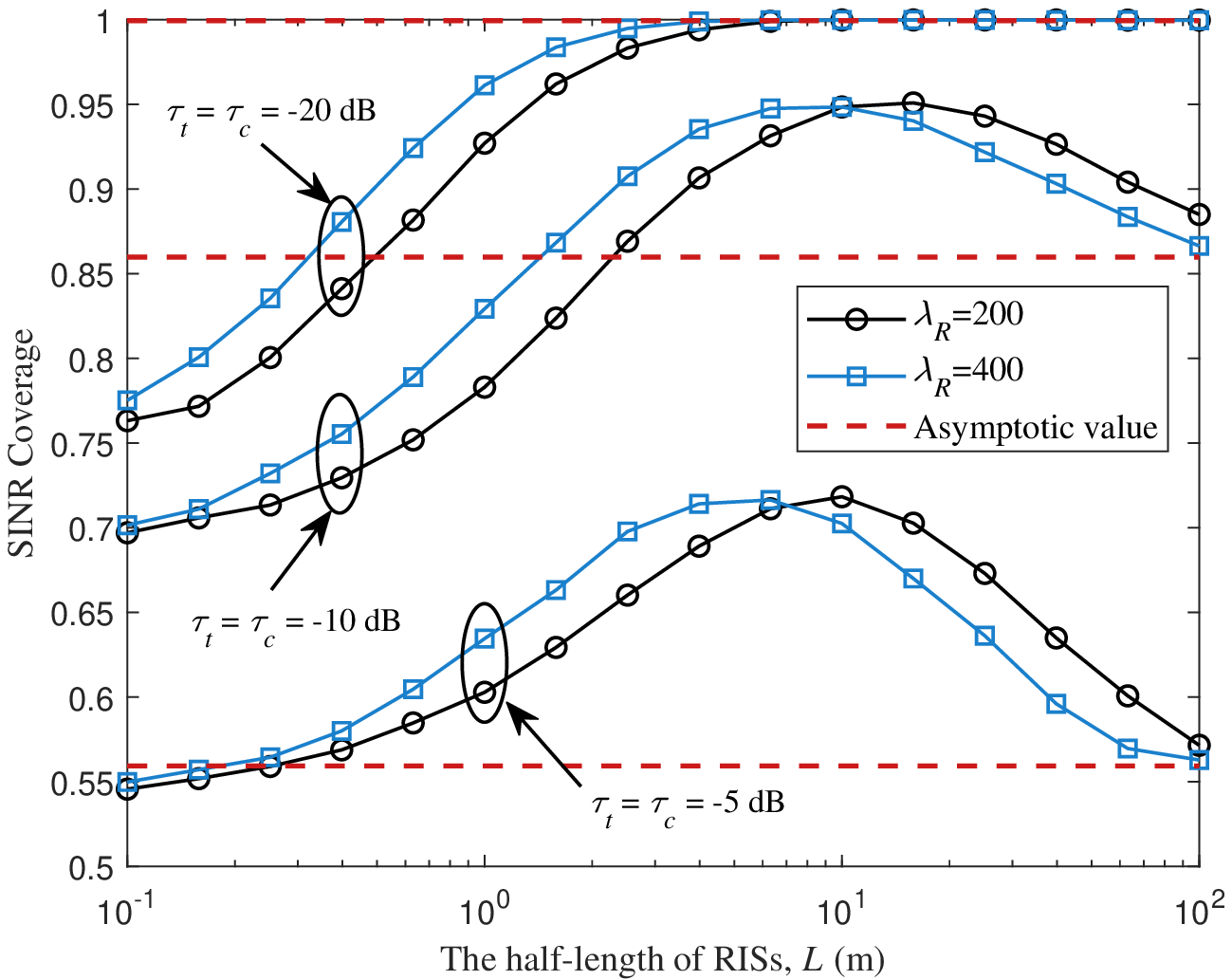}
	\caption{SINR coverage probability versus half-length of RISs with $\lambda_B = 10$ km$^{-2}$ and $P_B = 1$ W.
	}
	\label{figure_SINR_length}
\end{figure}

The SINR coverage probability comparison among NOMA RIS HetNets, OMA RIS HetNets and NOMA macro cells scenarios are shown in Fig. \ref{figure_SINR_comparison}. We can observe that comparing to macro cells scenarios, introducing the RIS HetNet structure enhances the SINR coverage performance considerably. This is attributed to the fact that RISs help the blocked UEs associate to NLoS BSs by a LoS reflective transmission path which has much smaller loss. Besides, RISs also supplement SINR coverage of LoS BSs.

\begin{figure} [t!]
	\centering
	\includegraphics[width = 3.7in] {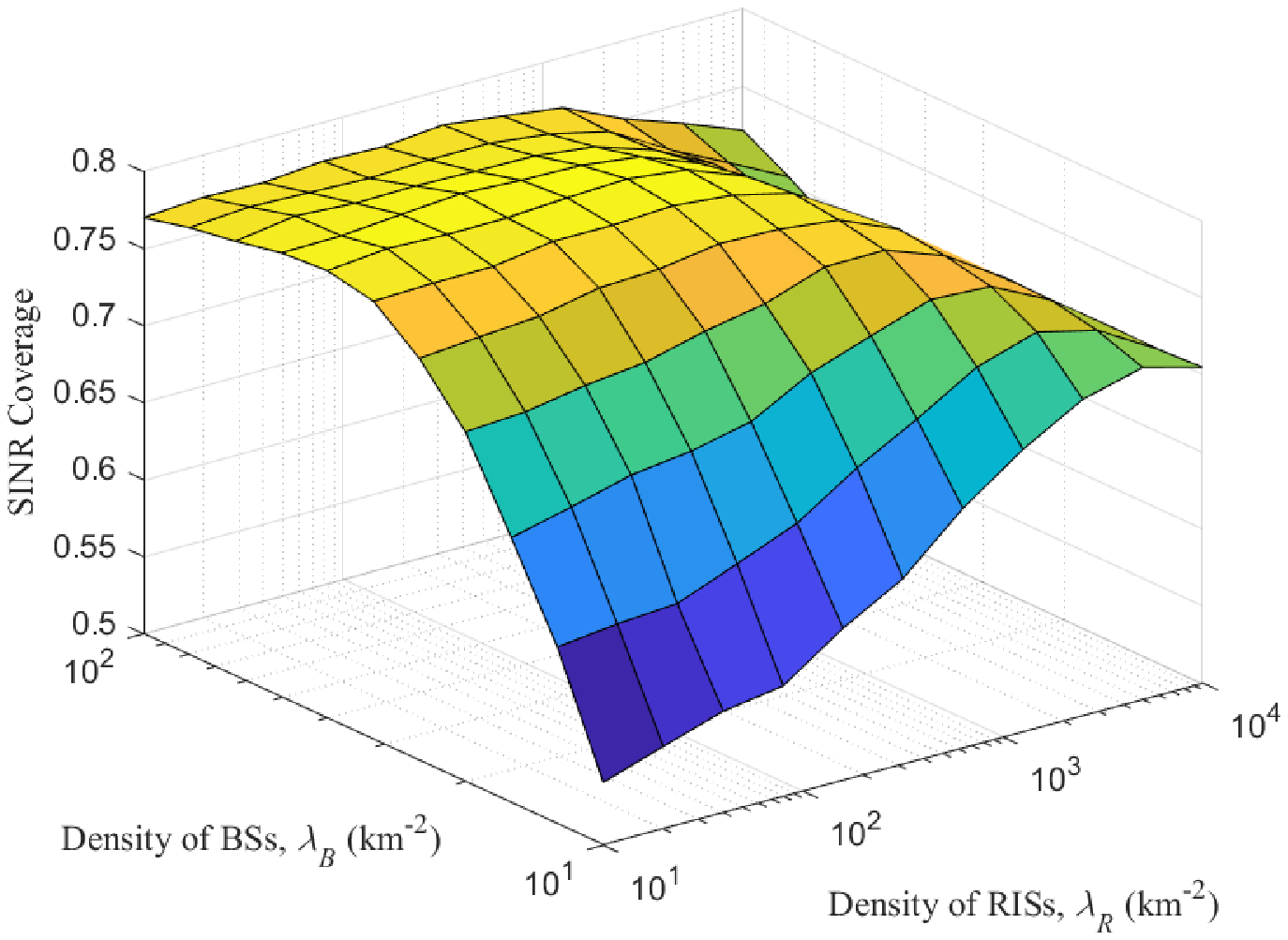}
	\caption{SINR coverage probability versus density of BSs and density of RISs with $\tau_t=\tau_c = -5$ dB and $P_B = 1$ W.
	}
	\label{figure_SINR_RIS_density}
\end{figure}
\begin{figure} [t!]
	\centering
	\includegraphics[width = 3.7in] {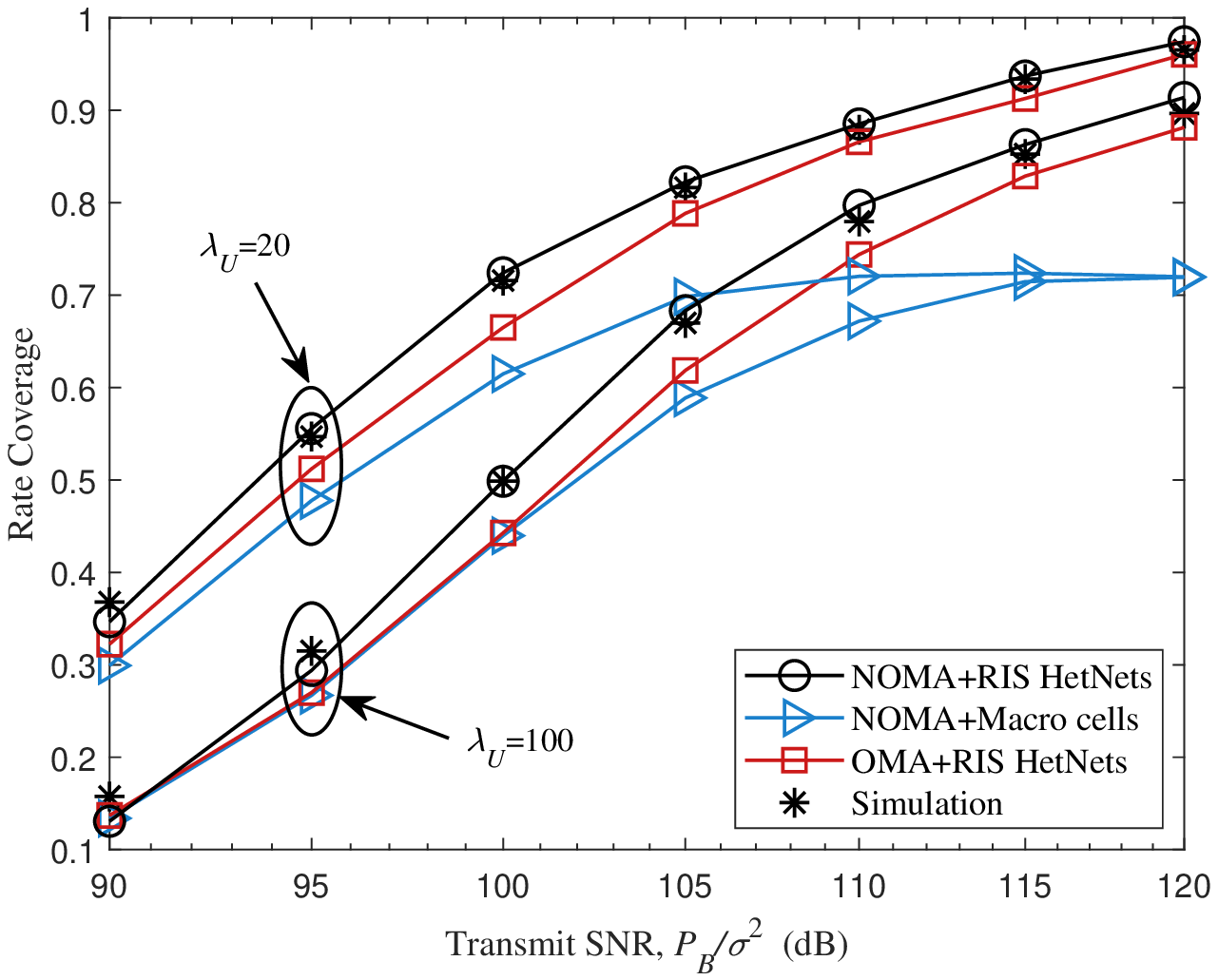}
	\caption{Rate coverage versus transmit SNR with $\lambda_U = [20,100]$ km$^{-2}$: a comparison among NOMA HetNets, OMA RIS HetNets and NOMA Macro cells.
	}
	\label{figure_rate_comparison}
\end{figure}

Fig. \ref{figure_SINR_length} plots the SINR coverage versus the half-length of RISs $L$ for $P_B=1$ W and $\lambda_B=10$ km$^{-2}$. We observe that there exists an optimal $L$ to maximize the system performance. For small-sized RISs, increasing the length of RISs improves the SINR coverage probability. This is because that as we increase the length, the channel gain of the BS-RIS-UE link grows and the interference power is dominated by both LoS BSs and RISs. However, with a further increase in the length of RISs, the signal as well as interference power from RISs grows rapidly. More UEs associate to RISs for large received power while these UEs suffer serious interference at the same time, which results in the decrease of SINR coverage. Finally, the value of SINR coverage asymptotically approaches a constant as discussed in \textbf{Corollary~\ref{corollary: Large length L}}.

Fig. \ref{figure_SINR_RIS_density} plots the SINR coverage versus the density of BSs $\lambda_B$ and the density of RISs $\lambda_R$. To observe the trend clearly, we set SINR threshold as $\tau_t = \tau_c = -5$ dB. Within the reasonable range of the parameters, there exists a optimal combination of $\lambda_B$ and $\lambda_R$ to maximize the SINR coverage. The reasons are two-fold: 1) similar to the trend of increasing the length of RISs, the SINR coverage decreases finally with the increase of the density of RISs due to the high interference power from RISs; 2) the densification of BSs brings stronger signal power as well as denser LoS interfering BSs.

\subsection{Rate Coverage}
In Fig. \ref{figure_rate_comparison}, we compare the rate coverage probability versus transmit SNR for NOMA RIS HetNets, OMA RIS HetNets and NOMA macro cells scenarios. Since the exact expression of the rate coverage in \textbf{Theorem~\ref{theorem: exact rate}} is intractable, the curves representing the performance of NOMA RIS HetNets are from \textbf{Corollary~\ref{corollary: approximate rate}}. It is observed that the rate coverage in NOMA RIS HetNets outperforms the counterpart in other two scenarios. This is becauese 1) the RIS HetNet structure brings higher channel gain than traditional macro cell networks; 2) the NOMA technique improves bandwidth efficiency. The enhancement of performance validates the effectiveness of our proposed NOMA HetNet framework.

\section{Conclusion}
In this paper, RIS-aided downlink NOMA networks with a HetNet structure have been investigated, where the stochastic geometry has been utilized for modeling the locations of BSs, RISs, and UEs and evaluating the system performance. A practical UE association scheme has been employed to maximize the average received power. Considering correlated RIS channel, the closed-form PDF of BS-RIS-UE links has been derived, based on which we have presented the UE association probability and its bounds.  The analytical expressions of SINR and rate coverage probability have been deduced and validated by numerical results. The coverage enhancement from NOMA schemes and the proposed RIS HetNet structure is verified by comparing to OMA RIS HetNets and NOMA macro cells scenarios. The analysis of this paper has provided guidance for deployment of RISs in the NOMA RIS NetNets: 1) there exists an optimal RIS size for maximum coverage performance; 2) when the density of BSs is known, deploying RISs with an appropriate density can maximize the system coverage.

\numberwithin{equation}{section}
\section*{Appendix~A: Proof of Lemma~\ref{lemma: PDF d_BU_RIS}}
\label{Appendix:A}
\renewcommand{\theequation}{A.\arabic{equation}}
\setcounter{equation}{0}
As $d_{0,j}^{(i)} = d_{BR,j}^{(i)}d_{RU,0}^{(i)}$, the CDF of $d_{0,j}^{(i)}$ can be given by
\begin{align}
    {F_{d_{0,j}^{(i)}}}(z)  &= \mathbb{P}(d_{BR,j}^{(i)}d_{RU,0}^{(i)} \le z)\nonumber
    \\& = \int_0^\infty  {\mathbb{P}(d_{BR,j}^{(i)} = x)\mathbb{P}(d_{RU,0}^{(i)} \le \frac{z}{x})dx} \nonumber
    \\& = \int_0^\infty  {{f_{d_{BR}}}(x)\int_0^{z/x} {{f_{d_{RU}}}(y)dy} dx}.
\end{align}

The PDF of $d_{0,j}^{(i)}$ can be obtained from the derivative of ${F_{d_{0,j}^{(i)}}}(z)$ as
\begin{align}\label{PDF d_BU_RIS2}
    {f_{d_{0,j}^{(i)}}}(z) = \frac{d}{{dz}}{F_{d_{0,j}^{(i)}}}(z) = \int_0^\infty  {\frac{1}{x}{f_{d_{BR}}}(x){f_{d_{RU}}}(z/x)dx}.
\end{align}

By substituting \eqref{PDF d_BR} and \eqref{PDF d_RU} into \eqref{PDF d_BU_RIS2}, the PDF of $d_{BU,j}^{(i)}$ can be expressed as
\begin{align}
    {f_{d_{0,j}^{(i)}}}(z) & = 4{\pi ^2}z{\lambda _B}{\lambda _R}\int_0^\infty  {\frac{1}{x}\exp \left( { - \pi {\lambda _B}{x^2} - \pi {\lambda _R}\frac{{{z^2}}}{{{x^2}}}} \right)dx} \nonumber\\
    &\overset{(a)}{=}  2{\pi ^2}z{\lambda _B}{\lambda _R}\int_0^\infty  {\frac{1}{t}\exp \left( { - \pi {\lambda _B}t - \pi {\lambda _R}\frac{{{z^2}}}{t}} \right)dt} \nonumber\\
    &\overset{(b)}{=} 4{\pi ^2}z{\lambda _B}{\lambda _R}{K_0}\left( {2\pi z\sqrt {{\lambda _B}{\lambda _R}} } \right),
\end{align}
where $(a)$ is obtained by employing the change of variable  $t=x^2$. $(b)$ is obtained by applying \cite[eq. (3.478.4)]{5}. The proof is completed.

\section*{Appendix~B: Proof of Lemma~\ref{Lemma: Interference LoS BSs}}
\label{Appendix:B}
\renewcommand{\theequation}{B.\arabic{equation}}
\setcounter{equation}{0}
The Laplace transform of the interference from LoS BSs can be expressed as follows
\begin{align}
    {{\cal L}_{{I_L}}}(s) &= {\mathbb{E}_{\Phi _B^L}}\left[ {\prod\limits_{k \in \Phi _B^L\backslash j} {{\mathbb{E}_{h{{_{0,k}^L}^2}}}\left[ {\exp \left( { - s{P_B}{C_L}{d_{0,k}}^{ - {\alpha _L}}h{{_{0,k}^L}^2}} \right)} \right]} } \right] \nonumber \\
    & \overset{(a)}{=} \exp \left( { - 2\pi {\tilde \lambda _B}\int_{{d_{0,min}}}^\infty  {\left( {1 - {\mathbb{E}_{h{{_{0,k}^L}^2}}}\left[ {\exp \left( { - s{P_B}{C_L}{x^{ - {\alpha _L}}}h{{_{0,k}^L}^2}} \right)} \right]} \right)x{p_L}(x)dx} } \right) \nonumber\\
    & \overset{(b)}{=} \exp \left( { - 2\pi {\tilde \lambda _B}\int_{{d_{0,min }}}^\infty  {\left( {1 - {{\left( {1 + \frac{{s{P_B}{C_L}}}{{{m_L}{x^{{\alpha _L}}}}}} \right)}^{ - {m_L}}}} \right)x{p_L}(x)dx} } \right),
\end{align}
where $(a)$ follows from probability generating functional (PGFL). $(b)$ is obtained by computing the moment generating function of the gamma random variable ${h{{_{0,k}^L}^2}}$.

Then we calculate the minimum interfering distance $d _{0,min}$. If the typical UE associates to a LoS BS, $d _{0,min}^L = d_0$. If the typical UE is aided with a RIS, the nearest interfering LoS BS $k \in \Phi _B^L$ satisfies
\begin{align}
    {C_L}{d_{0,k}}^{ - {\alpha _L}} \le {C_R}{d{_{0,j}^{(i)}}^{ - {\alpha _R}}}.
\end{align}

Therefore, $d _{0,min}^R = \min\; d_{0,k} ={( {{{\tilde c}_{LR}}})^{\frac{1}{{{\alpha _L}}}}}{d{_{0,j}^{(i)}}^{{{\tilde \alpha }_{RL}}}}$. According to the maximum average received power association scheme applied in this work, $d _{0,min}= \min\{d _{0,min}^L,d _{0,min}^R\}$. This lemma is proved.

\section*{Appendix~C: Proof of Lemma~\ref{Lemma: Interference RISs}}
\label{Appendix:C}
\renewcommand{\theequation}{C.\arabic{equation}}
\setcounter{equation}{0}
The Laplace transform of the interference from RISs can be expressed as
\begin{align}
    {{\cal L}_{{I_R}}}(s) &= {\mathbb{E}_{\Phi _B^R}}\left[ {\prod\limits_{k \in \Phi _B^R\backslash j} {{\mathbb{E}_{h{{_{0,k}^R}^2}}}\left[ {\exp \left( { - s{P_B}{C_R}{{\left( {d_{BR,k}^{(i)}d_{RU,0}^{(i)}} \right)}^{ - {\alpha _R}}}h{{_{0,k}^R}^2}} \right)} \right]} } \right] \nonumber \\
    &\overset{(a)}{=} \exp \left( { - \pi {\tilde \lambda _B}\int_{d_{BR,j}^{(i)}}^\infty  {\left( {1 - {\mathbb{E}_{h{{_{0,j}^R}^2}}}\left[ {\exp \left( { - s{P_B}{C_R}{x^{ - {\alpha _R}}}d{{_{RU,0}^{(i)}}^{ - {\alpha _R}}}h{{_{0,k}^R}^2}} \right)} \right]} \right)xdx} } \right) \nonumber \\
    &= \exp \left( { - \pi {\tilde \lambda _B}\int_{d_{BR,j}^{(i)}}^\infty  {\left( {1 - {{\left( {1 + \frac{{s{P_B}{C_R}}}{{{m_R}{{\left( {d_{RU,0}^{(i)}x} \right)}^{{\alpha _R}}}}}} \right)}^{ - {m_R}}}} \right)xdx} } \right),
\end{align}
where $(a)$ follows the fact that BSs located at the back of the RIS would not become the interfering BSs.

We denote $A = \frac{{s{P_B}{C_R}}}{{{m_R}d{{_{RU,0}^{(i)}}^{{\alpha _R}}}}}$, $B = d_{BR,j}^{(i)}$, and $C = \pi {\tilde \lambda _B}$ for simplicity. The Laplace transform can be rewritten as
\begin{align}
   {{\cal L}_{{I_R}}}(s) &= \exp \left( { - C\int_B^\infty  {\left( {1 - {{\left( {1 + A{x^{ - {\alpha _R}}}} \right)}^{ - {m_R}}}} \right)xdx} } \right) \nonumber \\
   &\overset{(b)}{=} \exp \left( {\frac{C}{{{\alpha _R}A}}\int_0^{ - A{B^{ - {\alpha _R}}}} {\left( {1 - {{\left( {1 - t} \right)}^{ - {m_R}}}} \right){{\left( { - \frac{t}{A}} \right)}^{ - 2/{\alpha _R} - 1}}dx} } \right) \nonumber \\
   &\overset{(c)}{=} \exp \left( {\frac{{C{B^2}}}{2} - \frac{{C{B^2}}}{2}{}_2{F_1}\left( {{m_R},-\frac{2}{{{\alpha _R}}};1 - \frac{2}{{{\alpha _R}}}; - A{B^{ - {\alpha _R}}}} \right)} \right)
\end{align}
where $(b)$ is obtained by employing the change of variable  $t=-Ax^{-\alpha_R}$. $(c)$ is obtained by applying \cite[eq. 8.391]{5}. The proof is completed.

\section*{Appendix~D: Proof of Theorem~ \ref{theorem: L/R UE cp}}
\label{Appendix:D}
\renewcommand{\theequation}{D.\arabic{equation}}
\setcounter{equation}{0}

Since there are two kinds of links for the typical UE to associates to a BS, let $T = \{ L,R \}$ denote the association choice of the typical UE, the probability that the SINR requirement of a random UE $u_0 \in \Phi _U$ is met is
\begin{align}
{P_{cov }}({\tau _c},{\tau _t}) =\sum\limits_{T = \{ L,R\} }{P_{cov }^{T}}({\tau _c},{\tau _t}) = \sum\limits_{T = \{ L,R\} } {\mathbb{P}\left( {\gamma _t^T > {\tau _t} \cap \gamma _{t \to c,small}^T > {\tau _c}},{{u_0} \in \Phi _U^T} \right)}.
\end{align}

For the case that the typical UE associates to a LoS BS directly, $P_{0,j}>P_{0,j}^{(i)}$ holds, so we have $d_{BR,j}^{(i)}d_{RU,0}^{(i)}>\varphi^{-1} (d_{0,j})$, where $\varphi^{-1} (x) = {\left( {{{\tilde c}_{RL}}} \right)^{\frac{1}{{{\alpha _R}}}}}{x^{{{\tilde \alpha }_{LR}}}}$. The SINR coverage probability can be expressed as
\begin{align}
&P_{cov}^L({\tau _c},{\tau _t}){\rm{ = }}\int_0^{{d_C}} {\int_{{\varphi ^{ - 1}}(x)}^\infty  {P_{cov,small}^L(x){f_{{d_{0,j}}}}(x)} } {f_{d_{0,j}^{(i)}}}(y)dydx\nonumber\\
&+ \int_{{d_C}}^\infty  {\int_{{\varphi ^{ - 1}}(x)}^\infty  {P_{cov,large}^L(x){f_{{d_{0,j}}}}(x)} } {f_{d_{0,j}^{(i)}}}(y)dydx.
\end{align}

For the case that the typical UE associates to a RIS, the association condition can be rewritten as ${d_{0,j}} > \varphi(d_{BR,j}^{(i)}d_{RU,0}^{(i)})$, so the SINR coverage probability can be expressed as
\begin{align}
&P_{cov}^R({\tau _c},{\tau _t}) = \int_0^\infty  {\int_0^{\zeta ({d_C})/{x_1}} {\int_{\varphi \left( {{x_1}{x_2}} \right)}^\infty  {P_{cov,small}^R\left( {{x_1},{x_2}} \right){f_{{d_{0,j}}}}\left( {{x_3}} \right)d{x_3}} {f_{{d_{BR}}}}\left( {{x_2}} \right)d{x_2}} } {f_{{d_{RU}}}}\left( {{x_1}} \right)d{x_1}\nonumber\\
&+ \int_0^\infty  {\int_{\zeta ({d_C})/{x_1}}^\infty  {\int_{\varphi \left( {{x_1}{x_2}} \right)}^\infty  {P_{cov,large}^R\left( {{x_1},{x_2}} \right){f_{{d_{0,j}}}}\left( {{x_3}} \right)d{x_3}} {f_{{d_{BR}}}}\left( {{x_2}} \right)d{x_2}} } {f_{{d_{RU}}}}\left( {{x_1}} \right)d{x_1}.
\end{align}

Noticed that the integral $\int_A^\infty  {{f_{{d_{0,j}}}}\left( x \right)} dx = 1 - {F_{{d_{0,j}}}}(A) = {{\bar F}_{{d_{0,j}}}}(A)$, \eqref{eq: cp for RIS} is obtained. The proof is completed.

\section*{Appendix~E: Proof of Corollary~\ref{corollary: Large length L}}
\label{Appendix:E}
\renewcommand{\theequation}{E.\arabic{equation}}
\setcounter{equation}{0}

When $L \to \infty$, the intercept $C_R \to \infty$, hencee the path loss for the BS-RIS-UE link is always smaller than the BS-UE link. Therefore, the typical UE always associates to the RIS, i.e. $A_R = 1$ holds. In this case, the SINR coverage probability for the typical UE is rewritten as $P_{cov}({\tau _c},{\tau _t})=P_{cov}^R({\tau _c},{\tau _t})$. Besides, for a finite value $x$, $\varphi(x) \to 0$ holds. The SINR coverage can be expressed as
\begin{align}
    &P_{cov}({\tau _c},{\tau _t}) =
    \int_0^\infty  {\int_0^{\zeta ({d_C})/{x_1}} {P_{cov,small}^R\left( {{x_1},{x_2}} \right){f_{{d_{BR}}}}\left( {{x_2}} \right)d{x_2}} } {f_{{d_{RU}}}}\left( {{x_1}} \right)d{x_1} \nonumber\\
    & + \int_0^\infty  {\int_{\zeta ({d_C})/{x_1}}^\infty  {P_{cov,large}^R\left( {{x_1},{x_2}} \right){f_{{d_{BR}}}}\left( {{x_2}} \right)d{x_2}} } {f_{{d_{RU}}}}\left( {{x_1}} \right)d{x_1}\nonumber \\
    &\overset{(a)}{=} \int_0^\infty  {\int_0^\infty  {P_{cov,small}^R\left( {{x_1},{x_2}} \right){f_{{d_{BR}}}}\left( {{x_2}} \right)d{x_2}} } {f_{{d_{RU}}}}\left( {{x_1}} \right)d{x_1}\nonumber \\
    &\overset{(b)}{\approx} \int_0^\infty  {\int_0^\infty  {\sum\limits_{n = 1}^{{m_R}} {{{( - 1)}^{n + 1}}} {\binom{m_R}{n}} {{\cal L}_{{I_R}}}({s_R}){f_{{d_{BR}}}}\left( {{x_2}} \right)d{x_2}} } {f_{{d_{RU}}}}\left( {{x_1}} \right)d{x_1} \nonumber \\
    & = 2\pi {\lambda _B}\sum\limits_{n = 1}^{{m_R}} {{{( - 1)}^{n + 1}}} {\binom{m_R}{n}}\int_0^\infty  {{x_2}} \exp \left( { - \pi \left( {{{\tilde \lambda }_B}A + {\lambda _B}} \right){x_2}^2} \right)d{x_2}\nonumber \\
    & = \sum\limits_{n = 1}^{{m_R}} {{{( - 1)}^{n + 1}}} {\binom{m_R}{n}}\frac{{{\lambda _B}}}{{{{\tilde \lambda }_B}A + {\lambda _B}}}
    ,
\end{align}
where $A = \frac{1}{2}{Q_R}\left( {{\eta _R},{\tau ^*}} \right)$ and ${{Q_R}\left( {\eta ,\tau } \right)}= {}_2{F_1}\left( {{m_R}, - \frac{2}{{{\alpha _R}}};1 - \frac{2}{{{\alpha _R}}}; - \frac{{n\eta \tau }}{{{m_R}}}} \right) - 1$. $(a)$ is resulted from $\zeta ({d_C}) \to \infty$. $(b)$ is obtained by using the fact that the inteference power from RISs dominates the aggregate interference and the noise power is negligible when $L \to \infty$ . The proof is completed.
\begin{spacing}{1.5}
\bibliographystyle{IEEEtran}
\bibliography{mybib}
\end{spacing}
\end{document}